\documentstyle[epsf,times]{mn}
\title[Properties of the XBACs --- I. The sample]
      {Properties of the X-ray brightest Abell-type clusters of galaxies (XBACs)
       from ROSAT All-Sky Survey data --- I. The sample}
\author[Ebeling et al.]
       {\parbox{\textwidth}{H.\ Ebeling$^{1,2,\star}$,
        W.\ Voges$^1$, H.\ B\"ohringer$^1$, A.C.\ Edge$^2$, J.P.\ Huchra$^3$, 
        U.G.\ Briel$^1$}\\ \\
        $^1$ Max-Planck-Institut f\"ur extraterrestrische Physik, 
             Giessenbachstr., D-85740 Garching, Germany\\
 	$^2$ Institute of Astronomy, Madingley Road, Cambridge CB3\,0HA, UK\\
	$^3$ Harvard-Smithsonian Center for Astrophysics, 60 Garden Street, 
             Cambridge, MA 02138, USA \\
        $^\star$ send correspondence to ebeling{@}ast.cam.ac.uk or ebeling{@}ifa.hawaii.edu}

\date{Received ***; in original form 1995 September ***}
\begin{document}

\maketitle

\begin{abstract} 

We present an essentially complete, all-sky, X-ray flux limited sample
of 242 Abell clusters of galaxies (six of which are double) compiled
from ROSAT All-Sky Survey data. Our sample is uncontaminated in the
sense that systems featuring prominent X-ray point sources such as AGN
or foreground stars have been removed. The sample is limited to high
Galactic latitudes ($|b| \geq 20^{\circ}$), the nominal redshift range
of the ACO catalogue of $z \leq 0.2$, and X-ray fluxes above $5.0
\times 10^{-12}$ erg cm$^{-2}$ s$^{-1}$ in the 0.1 -- 2.4 keV band.
Due to the X-ray flux limit, our sample consists, at intermediate and
high redshifts, exclusively of very X-ray luminous clusters.  Since
the latter tend to be also optically rich, the sample is not affected
by the optical selection effects and in particular not by the volume
incompleteness known to be present in the Abell and ACO catalogues for
richness class 0 and 1 clusters.

Our sample is the largest X-ray flux limited sample of galaxy clusters
compiled to date and will allow investigations of unprecedented
statistical quality into the properties and distribution of rich
clusters in the local Universe.

\end{abstract}


\section{Introduction} 

Being the largest gravitationally bound systems to have decoupled from
the Hubble expansion of the Universe, clusters of galaxies are ideal
tracers of the formation and evolution of structure on the largest
mass scales.  In the past, galaxy clusters were detected and
classified mainly on the grounds of their optical appearance, the
largest compilations of this kind being the catalogues of Abell
(1958), Abell, Corwin \& Olowin (1989, from here on ACO), and Zwicky
and co-workers (1961--1968). In particular the Abell and ACO
catalogues have been used extensively for statistical studies of
cluster properties, but served also as finding lists for more detailed
investigations of individual clusters.

The major drawback of optical compilations of this kind lies in the
high risk of projection effects corrupting the sample. Since the
cluster detection and selection is performed on the two-dimensional
distribution of galaxies as it appears in projection on optical
plates, fluctuations in the surface density of field galaxies as well
as superpositions of poor clusters along the line of sight can lead to
an overestimation of a system's richness.  On the other hand, poor
clusters can be missed completely as they often do not contrast
strongly with the background field. Several studies have investigated
the statistical completeness and contamination of the Abell catalogue
emphasizing the importance of projection effects for optically
selected cluster samples (Lucey 1983, Sutherland 1988, Struble \& Rood
1991).

The serious problem of projection effects can, however, be overcome
completely by selecting clusters in the X-ray rather than in the
optical. X-ray emission from a diffuse, gaseous intra-cluster medium
(ICM) trapped and heated to temperatures of typically a few $10^7$ K
is not only a sure indication that the system is indeed
three-dimensionally bound. Being caused by ion-ion collisions in the
ICM, the X-ray flux is proportional to the square of the ion density
and thus much more peaked at the gravitational centre of the cluster
than the projected galaxy distribution. This property requires
clusters to be almost perfectly aligned along the line of sight in
order to be mistaken for a single, more luminous entity, so that
projection effects caused by such superpositions can be effectively
neglected in the X-ray.

X-rays thus provide a very efficient and unbiased way of compiling
cluster samples which, as far as statistical studies are concerned,
will eventually supersede the present, optical catalogues. Early
statistical cluster samples have been compiled from the X-ray data
taken by the {\sc Uhuru} (Schwartz 1978), {\sc Ariel V} (McHardy
1978), and {\sc Heao 1 A-2} satellites (Piccinotti et al.\ 1982), all
of which comprised about 30 clusters. Including clusters detected
during the {\sc Exosat} mission, Lahav et al.\ (1989) and Edge et al.\
(1990) presented an X-ray flux limited sample of 55 clusters, very
similar in size to that compiled by Gioia et al.\ (1990) from X-ray
sources detected in the {\sc Einstein} Medium Sensitivity Survey.
Projects aimed at the compilation of 
purely X-ray selected cluster samples based on the X-ray data
collected during the ROSAT All-Sky Survey (RASS, Voges 1992) are well
under way in both the northern (Allen et al.\ 1992, Giacconi \& Burg
1993, Crawford et al.\ 1995) and the southern hemisphere (Guzzo et
al.\ 1995, De Grandi 1996, see also Pierre et al.\ 1994); the
two-point correlation function for RASS selected cluster samples
compiled in smaller sky areas has been presented recently (Nichol et
al.\ 1994, Romer et al.\ 1994) showing again the advantage of X-ray
over optical selection procedures.  An overview of the various RASS
cluster projects is given in B\"ohringer (1994).

In the following, we shall pursue an approach that investigates the
X-ray properties of optically selected clusters and leads to an
all-sky sample of the X-ray brightest Abell-type clusters of galaxies
(XBACs)\footnote{An early, if incomplete, version of this sample has
been presented by Ebeling (1993).}  which is more than four times
larger than the largest X-ray flux limited cluster sample published to
date.

In this first article of a series we present the sample and discuss
its statistical properties; in subsequent papers we shall establish
the X-ray luminosity function of Abell-type clusters of galaxies
(Ebeling et al., in preparation) as well as the cluster-cluster
correlation function (Edge et al., in preparation), and investigate
correlations between X-ray and optical properties of the XBACs.

Throughout this paper, we assume an Einstein-de Sitter Universe with
$q_0 = 0.5$ and $H_0 = 50$ km s$^{-1}$ Mpc$^{-1}$.

\section{The optical database}
\label{m10cal}

All optical Abell cluster parameters used here, including the cluster
coordinates, are taken from ACO's 1989 publication. Of the original
4076 clusters in the catalogue, we discard A\,3541 and A\,3897 as
secondary detections of A\,1664 and A\,2462, respectively.

Besides the actual cluster coordinates the most crucial cluster
parameter for both the compilation of our sample and the subsequent
analysis of various physical cluster properties is possibly the
cluster's redshift. Although the number of measured redshifts for
Abell and ACO clusters of galaxies has seen an impressive increase in
the past decade, we still have to rely on estimated redshifts for the
large majority of the more distant clusters. In the following
paragraphs we shall describe briefly the procedure adopted by us for
the estimation of redshifts for ACO clusters without, or with
questionable, measured redshifts.

Attempts to estimate redshifts from different optical parameters, such
as $m_1$, $m_3$, $m_{10}$, the apparent cluster diameter or a
combination of any of these, have been described by various authors
(Abell 1958, Corwin 1974, Leir \& van den Bergh 1977, Postman et al.\
1985, ACO, Couchman et al.\ 1989, Scaramella et al.\ 1991, Peacock \&
West 1992). Our calibration relies on the most commonly used redshift
estimator, $m_{10}$, which is statistically more robust than $m_1$ or
$m_3$ and has the additional advantage of being available for all of
the catalogued clusters.

\begin{figure}
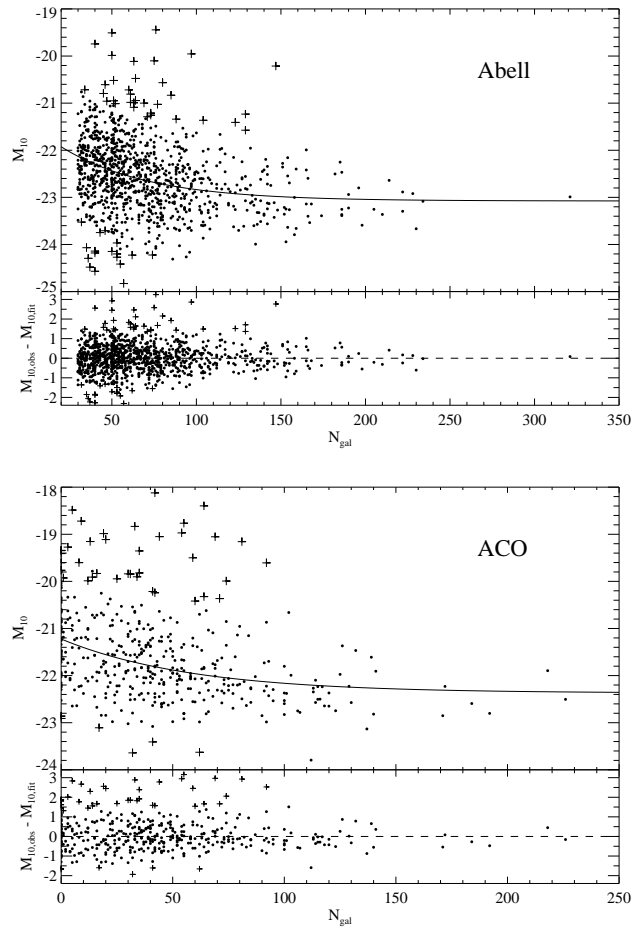

	\epsfxsize=0.5\textwidth
	\hspace*{0cm} \centerline{\epsffile{scott_eff_north.epsf}}
	\epsfxsize=0.5\textwidth
	\hspace*{0cm} \centerline{\epsffile{scott_eff_south.epsf}}
	\caption[]{The richness dependence of $M_{10}$, the Scott 
                 effect, for Abell and ACO clusters. For each sample the 
                 least-squares fit of a constant plus an exponential is 
	         overlaid; the bottom 
	         panels show the respective residuals. Clusters marked by
                 crosses are those discarded as inconsistent with the
	         $m_{10}$-$z$ relation.}
	\label{scott_eff}
\end{figure}
The $m_{10}$-$z$ relation used in the following to obtain distance
estimates for those clusters which do not have measured $z$'s is
calibrated with 1\,482 Abell and ACO clusters with known redshifts
(including 182 southern supplementary ACO clusters), the great
majority of which have been compiled by one of us (JPH). Additional
ACO cluster redshifts are taken from Heinz Andernach's compilation
(Andernach 1991) and the NASA Extragalactic Database NED. Also, we are
indebted to C.A.\ Collins, A.K.\ Romer, C.S.\ Crawford and G.B.\
Dalton for allowing us to use, prior to publication, some of the
redshifts taken by them for different projects.  As for the cluster
redshifts obtained in the APM redshift survey, we use the Abell
identifications and redshifts given by Ebeling \& Maddox (1995) rather
than those published by Dalton et al.\ (1994).

Note that, although we do take the ACO supplementary clusters into
account for the purpose of the $m_{10}$-$z$ calibration, we do not
consider them in the rest of this paper.

For the actual calibration, the `raw' $m_{10}$ values in the
respective photometric system (R in the north, V in the south) are
corrected only for Galactic extinction. In the north, the values
listed by Abell can thus be used straight away, whereas in the south
ACO's V magnitudes are corrected for Galactic extinction following the
procedure described by Fisher \& Tully (1981).

The K-correction is taken to be $K(z) = 1.122\, z$ for the northern
red plates (Postman et al.\ 1985) and $K(z) = 4.14\, z-0.44\,z^2$ for
the southern IIIa-J plates\footnote{Strictly speaking, these
K-corrections are valid only for elliptical and S0 galaxies; assuming
that the tenth brightest galaxy is of that type does not introduce a
big error, however, as spirals are rare among the bright galaxies in
rich clusters.} (Ellis 1983).  For the K-corrected magnitudes the
correction for the Scott effect (Scott 1957) is then determined from
the dependence of the absolute magnitude of the tenth brightest
cluster galaxy, $M_{10}$, on the cluster richness as given by the
Abell number of galaxies, $n_{\rm gal}$. We model the Scott effect as an
exponential plus a constant,
\begin{equation}
     M_{10} (n_{\rm gal}) = c_0 + c_1\,\exp (c_2\,n_{\rm gal}),
 \label{scott_eqn}
\end{equation}
where the coefficients $c_i$ are obtained in separate fits to the data
in the northern and southern hemisphere. The correction term is then
given by $\Delta M_{\rm 10,Scott} = M_{10}(n_{\rm gal}) - 
M_{10}(50)$, i.e., a nominal richness of $n_{\rm gal}=50$ is used as
a reference point. Applying the corrections for
both reddening and the Scott effect then yields a tentative
$m_{10}$-$z$ relation which we take to be of the simple form
$\log_{10} z_{\rm est} = a + b\;m_{10}$.  The final coefficients $c_i$
(listed in Table~\ref{scott_tab}) as well as $a$ and $b$ are
determined in an iteration loop where redshifts with $(z_{\rm
obs}-z_{\rm est})/z_{\rm obs} \geq 3 \sigma$ or $(z_{\rm obs}-z_{\rm
est})/z_{\rm est} \geq 3 \sigma$ are progressively removed. This
approach leads to the exclusion of 52 northern Abell clusters (leaving
984) and 48 southern ACO main and supplementary clusters (leaving 398)
from the calibration.

\begin{table}
 \begin{centering}  
  \begin{tabular}{cccc} \hline \\[-5pt]
                &   $c_0$  & $c_1$ &   $c_2$  \\[5pt] \hline \\[-5pt]
	north	& $-23.08$ & 1.70  & $-0.020$ \\
	south	& $-22.37$ & 1.15  & $-0.017$ \\[5pt] \hline
  \end{tabular}
  \caption[]{The final values for the Scott--effect coefficients of
           Eq.~\protect\ref{scott_eqn} obtained in least--squares fits to the
           richness-$M_{10}$ distributions of
	   Abell clusters (north) and ACO clusters (south).}
  \label{scott_tab} 
 \end{centering}  
\end{table}

\begin{figure}
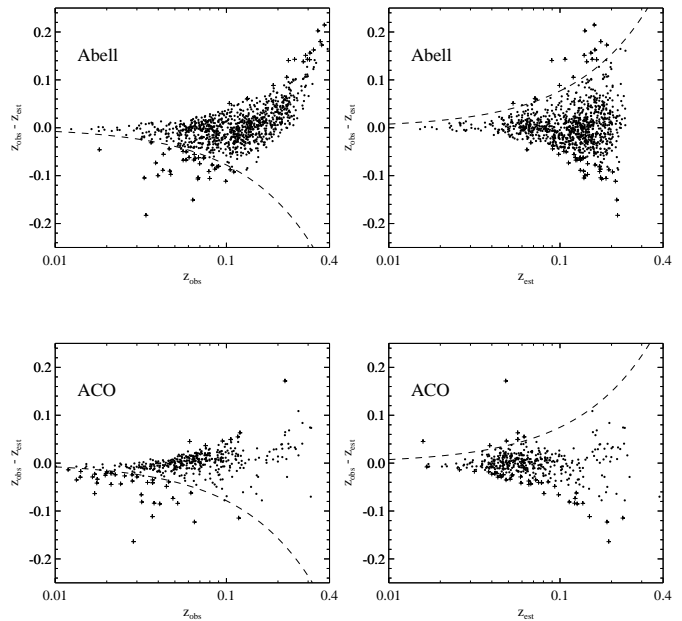

	\epsfxsize=0.5\textwidth
	\epsffile{z_delz_north.epsf}
	\epsfxsize=0.5\textwidth
	\epsffile{z_delz_south.epsf}
	\caption[]{The difference between measured and estimated redshifts as
                 a function of $z$ and $m_{10}$, respectively, for Abell and
		 ACO clusters. For each sample the 3$\sigma$ limits from the 
		 $m_{10}$-$z$ relation are overlaid as dashed lines. 
                 Clusters marked by
                 crosses are those discarded as inconsistent with the
	         $m_{10}$-$z$ relation.}                   
	\label{z_delz}
\end{figure}

\begin{figure}
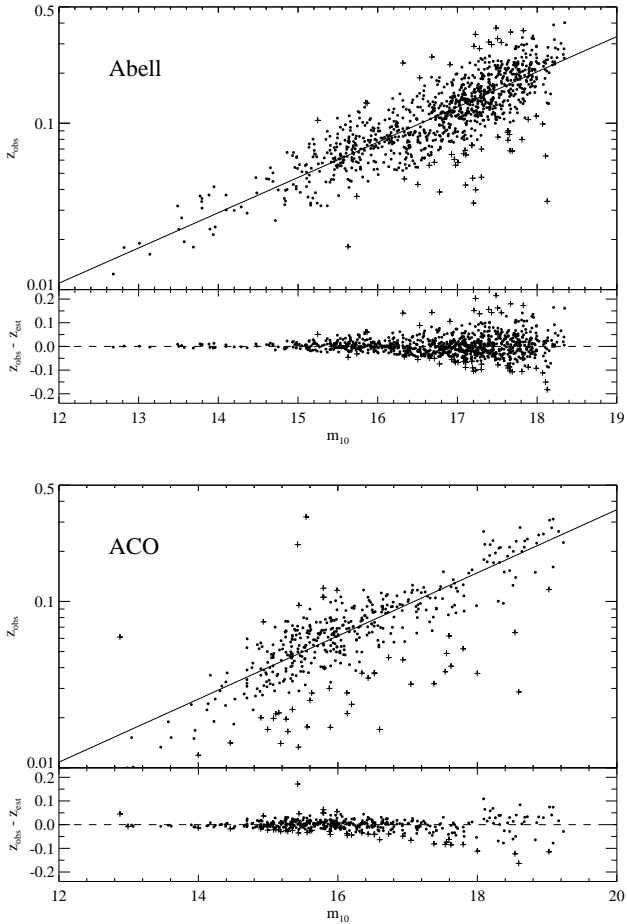

	\epsfxsize=0.5\textwidth
	\epsffile{m10_z_north.epsf}
	\epsfxsize=0.5\textwidth
	\epsffile{m10_z_south.epsf}
	 \caption[]{The Hubble diagrams for 984 Abell and 398 ACO main and
                 supplementary clusters. The solid lines represent the 
	 	 least-squares fits to the data; for each sample the bottom 
                 panel shows the respective residuals on a linear scale.
		 Clusters marked by crosses are those discarded as inconsistent 
		 with the $m_{10}$-$z$ relation.}                   
	\label{hubble}
\end{figure}

Figures~\ref{scott_eff} to \ref{hubble} show the resulting
distributions for Abell and ACO clusters as well as the residuals from
the respective least-squares fits.

The $m_{10}$-$z$ relation finally adopted by us is then given by
\begin{eqnarray*}
  \log z_{\rm est} & = & (-4.50 \pm 0.56) + (0.212 \pm 0.033)\times\\
                   &   &(m_{10}-\Delta M_{\rm 10,Scott} - K(z_{\rm est}))  \\
               &   & \mbox{\hfill Abell clusters} \\
               & = & (-4.24 \pm 0.66) + (0.190 \pm 0.041)\times\\
               &   & (m_{10}-\Delta M_{\rm 10,Scott} - K(z_{\rm est})) \\
               &   & \mbox{\hfill ACO clusters} \nonumber 
\end{eqnarray*}
Note, firstly, that, just as in the calibration where the cluster
redshifts are known from measurements, the K-correction is computed
iteratively at the {\em estimated}\/ redshift, and, secondly, that for
both samples the slope of the linear calibration is in good agreement
with the theoretically expected value of 0.2.

For Abell clusters the $m_{10}$-$z$ relation thus obtained is
consistent with the findings of Postman et al.\ (1985) but has a lower
dispersion of $\sigma(\log z - \log z_{\rm est}) = 0.11$ (compared to
0.14).  For the southern ACO main and supplementary clusters we find
the same value of $\sigma(\log z - \log z_{\rm est}) = 0.11$. In both
cases the above mentioned 3$\sigma$ outliers have been ignored.

\begin{figure}
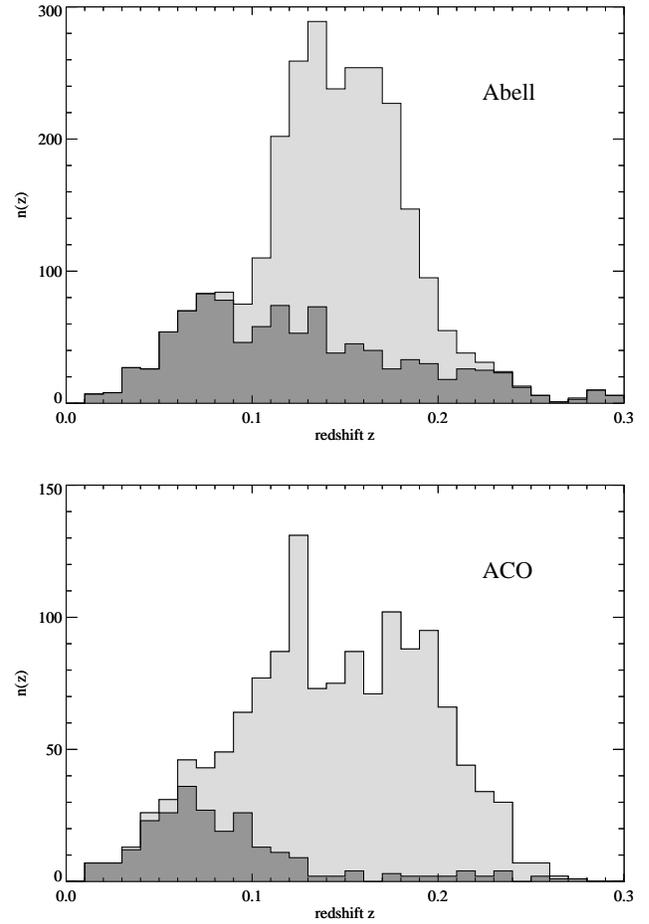

	\epsfxsize=0.5\textwidth
	\hspace*{0cm} \centerline{\epsffile{abe_zdist.epsf}}
	\epsfxsize=0.5\textwidth
	\hspace*{0cm} \centerline{\epsffile{aco_zdist.epsf}}
	\caption[]{The redshift distribution of Abell and ACO clusters of
		 galaxies combining measured (dark shading) and estimated 
                 values (light shading)}
	\label{aco_z_dist}
\end{figure}

Figure~\ref{aco_z_dist} finally shows the resulting redshift
distribution for Abell and ACO clusters. We should like to emphasize
once more that the southern supplementary clusters have been included
only temporarily for the purpose of our $m_{10}$-$z$ calibration but
are not included in the ACO sample as we use it here and in the
following. Also, the measured redshifts of clusters excluded from the
$m_{10}$-$z$ calibration because of inconsistencies between the
measured and estimated redshifts are not necessarily rejected as
interlopers, but scrutinized and only replaced by estimates if the
measured redshift is based on a single galaxy redshift that is clearly
in the foreground of the actual cluster. Having replaced 22 northern
and 30 southern cluster redshifts with estimates, we are left with
measured (and credible) redshifts for 1014 clusters from Abell's
original compilation and 246 southern ACO clusters.

\section{The X-ray database}
\label{rass_db}

From August 1990 to February 1991, the ROSAT X-ray satellite conducted
an all-sky survey (RASS) in the soft X-ray energy band ranging from
0.1 to 2.4 keV. Overviews of the ROSAT mission in general and the RASS
in particular can be found in the literature (Tr\"umper 1993, Voges
1992). A first processing of the data taken by the Position Sensitive
Proportional Counter (PSPC, Pfeffermann et al.\ 1986) during the RASS
was performed as the survey proceeded using the Standard Analysis
Software System (SASS) developed for this purpose (Voges et al.\ 1992)
at MPE. To this end, the incoming X-ray data were sorted into
two-degree wide strips of constant ecliptic longitude following the
satellite's scanning motion on the sky. Only two days worth of data
are collected in one strip, though, so that the exposure time in each
strip is roughly constant at some 360 s, and the much longer exposure
times accumulated around the ecliptic poles are not taken advantage
of.

\begin{figure}
  \epsfxsize=0.5\textwidth
  \hspace*{0cm} \centerline{\epsffile{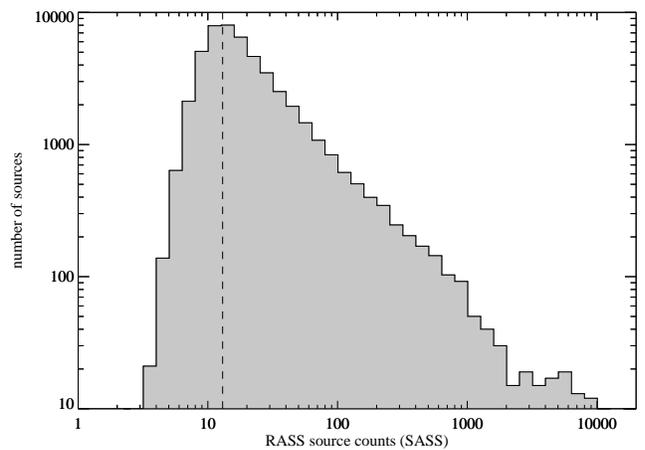}}
  \caption[]{The differential frequency distribution of the X-ray photon counts 
	   in RASS sources as detected by the SASS. Accordingly, the SASS 
	   master source list starts to become incomplete at source strengths 
	   of about 13 photons.}
  \label{sass_srccnts}
\end{figure}

Running on the ninety strips representing the RASS in this framework,
the SASS detected 49\,441 X-ray sources (multiple detections removed)
which constitute the X-ray sample most of the statistical RASS studies
performed to date are based upon. The source count distribution for
all 49\,441 SASS sources from this master source list is depicted in
Fig.~\ref{sass_srccnts} and shows that a minimum of some 10 photons is
required for any source in order to be detected reliably in this first
processing of the RASS data.

Since 1992 so-called PET files (for Photon Event Table) are produced
at MPE, which contain the full photon information in a field of
specified size around any given position in the sky thereby overcoming
the limitations of the strip data used in the earlier processing.

\section{A tentative sample}
\label{ten_sample}

\begin{figure}
  \epsfxsize=0.5\textwidth
  \hspace*{0cm} \centerline{\epsffile{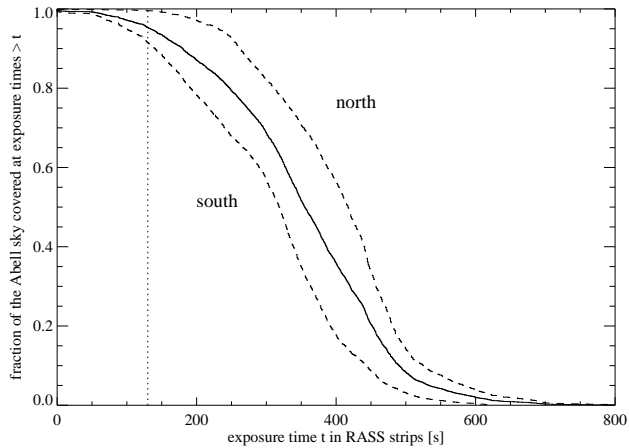}}
  \caption[]{The cumulative frequency distribution of the RASS exposure times
           in the strips the first SASS processing was performed on.
           Exposure times are computed in squares of 23 arcmin$^2$ area around
 	   the nominal positions of the 4074 catalogued ACO clusters. If a
           position falls into more than one strip, the highest exposure
	   time is used.}
  \label{rass_skycov}
\end{figure}

Starting from the results of the first SASS processing mentioned
above, we select a subset of 10\,241 sources with SASS count rates in
excess of 0.1 s$^{-1}$. This threshold is the result of a compromise
between maximal sky coverage and maximal survey depth: for sources
consisting of less than 13 photons the SASS source list starts to
become incomplete (see Fig.~\ref{sass_srccnts}), and at minimal
exposure times higher than 130 seconds the sky coverage of the RASS at
the positions of ACO clusters starts to fall below 95 per cent. The
latter figure is derived from the distribution shown in
Fig.~\ref{rass_skycov}. Note that in the southern hemisphere the
completeness with respect to Abell sky coverage is only 91.6 per cent
as opposed to 99.6 per cent in the north, due to the fact that, in
order to avoid damage to the detector, the PSPC is switched off
automatically when passing through regions of enhanced particle
background, for instance around the so-called South-Atlantic Anomaly.

\begin{figure}
  \epsfxsize=0.5\textwidth
  \hspace*{0cm} \centerline{\epsffile{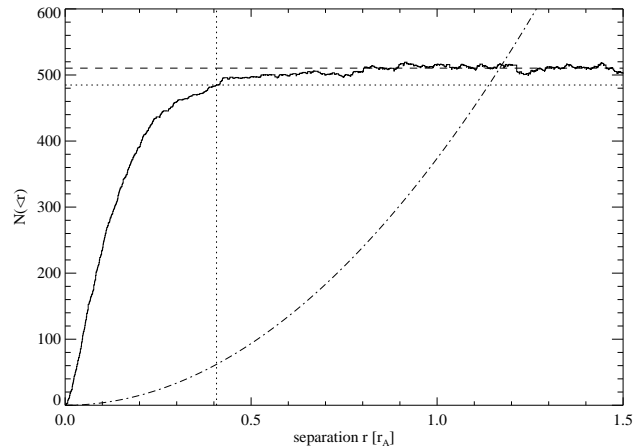}}
  \caption[]{The cumulative number of coincidences in the cross-correlation 
	   between the ACO catalogue and the RASS X-ray source list provided
	   by the SASS as a function of the X-ray to optical source 
           separation in Abell radii ($r_{\rm A}$). A parabolic background 
           component (the dot-dashed line)
           has been subtracted.}
  \label{sass_cumnum}
\end{figure}

We cross-correlate the reduced SASS X-ray source list with the ACO
cluster catalogue using the clusters' redshifts to scale angular to
metric separations. Estimated redshifts based on $m_{10}$ and taking
into account the cluster richness are used where measured redshifts
are not available (see Section~\ref{m10cal}). Figure~\ref{sass_cumnum}
shows the resulting cumulative distribution of the separations between
the SASS X-ray sources' and the optical cluster positions in units of
Abell radii ($r_{\rm A}$). A parabolic contribution from random
coincidences fitted to the data in the $0.75 \leq r/r_{\rm A} \leq
1.5$ range has been subtracted (the dash-dotted curve in
Fig.~\ref{sass_cumnum}). The total number of true coincidences of 510
is indicated by the dashed line in Fig.~\ref{sass_cumnum}.

We extract a 95 per cent complete sample by selecting the 547
correlation pairs (corresponding to 480 clusters) found out to
respective separations of 0.407 Abell radii. As even at $z=0.4$ this
cutoff still translates into an angular separation of more than 3
arcmin, it is always larger than the uncertainty in the optical
cluster positions of typically 2 to 3 arcmin (ACO) which therefore do
not need to be taken into account separately. Some 62 entries from
this list, i.e. 11 per cent, can be expected to be coincidental.

Of the 121 catalogued ACO clusters with $z \leq 0.05$ only 60 are
contained in this list. The lack of SASS detections in excess of 0.1
s$^{-1}$ for the remaining 61 nearby ACO clusters is, however, not
necessarily indicative of their being intrinsically X-ray
faint. Rather, it is possible that the SASS detection algorithms have
underestimated the flux from such potentially highly extended sources
or may have missed the source altogether, as the SASS was
designed solely for the detection of point sources.  We therefore also
include the remaining 61 ACO clusters with $z \leq 0.05$ which brings
the total number of ACO clusters in our tentative sample to 541.

For 535 of these, we obtained PET data in $2\times 2$ deg$^2$ fields
around the optical cluster position. As the PET file for any field
contains all photons accumulated during the whole RASS in the
respective region, the accumulated exposure times in these fields are
higher than those available in the SASS strips; up to 5800 s are
attained.  As mentioned before, the SASS has difficulties
characterizing or, for that matter, even reliably detecting extended
emission. The cause for this lies in the algorithm's basic design as a
point detection algorithm which is primarily sensitive to local and
quasi-spherical intensity variations. We therefore re-processed the
RASS data in the 535 fields with VTP (for Voronoi Tesselation \&
Percolation), an algorithm developed for the detection and
characterization of sources of essentially arbitrary shape (Ebeling \&
Wiedenmann 1993, Ebeling 1993).  VTP works on the individual photons
rather than on spatially binned representations of the data; also it
does not require any model profiles to be fitted to the detected
emission in order to determine fluxes all of which makes the algorithm
extremely versatile and flexible. The reprocessing of the data with
VTP is crucial in order to obtain reliable fluxes for the clusters in
our sample: for clusters of galaxies, raw SASS fluxes are found to be
typically too low by about a factor of 0.5; for nearby clusters the
misassessment can even exceed one order of magnitude (cf.\
Section~\ref{x_incomp}; see also Ebeling 1993).

Of the six potential cluster detections for which no PET files are
available, two can in fact be identified with stars whose positions
agree within a few arcsecs with those of the respective SASS source. A
third candidate features a spectral hardness ratio (see below) of
$-0.26$ and is also clearly off the cluster core as seen on the
optical plates, so that we can safely discard it as a non-cluster
source. Of the remaining three possible cluster detections, two
(A\,295 and A\,3089) were not detected by SASS at count rates greater
than 0.1 s$^{-1}$ and came in only because of their low redshift; in
fact A\,3089 is not detected by the SASS at any flux limit. Excluding
A\,3089, we are left with two detections, the second of which
(A\,3899) features an X-ray to optical separation of 0.4 $r_{\rm A}$
and, accordingly, has a very high probability of in fact being a
chance coincidence. We conclude that of the six clusters for which PET
data are not available for a VTP re-analysis at most two, namely
A\,295 and A\,3899, may qualify for inclusion in the XBACs sample.

The 3498 sources detected by VTP in the 535 PET fields were then
screened for possible blends of close source pairs. 228 potential
blends were visually inspected, and in 45 cases blended sources were
separated by introducing a dividing line into the
field. Figure~\ref{vtp_blend} illustrates the problem of VTP-produced
blends by showing the raw photon distribution and the VTP sources
detected around A\,2201.

\begin{figure*}
    \parbox{0.49\textwidth}{
    \epsfxsize=0.49\textwidth
    \epsffile{blend_a2201_p.epsf}}
    \parbox{0.49\textwidth}{
    \epsfxsize=0.49\textwidth
    \epsffile{blend_a2201_s.epsf}}
    \caption[]{A\,2201 as an example of a blended source. The plot
	     on the left shows the photon distribution in the centre of the
	     respective field; in the plot on the right the photons 
 	     constituting the VTP sources are overlaid as bold dots. The 
	     dotted lines indicate where VTP's percolation was forced to 
             stop in order to keep the two sources separated. The blended 
	     sources are A\,2201 in the field centre and, some 10 arcmins 
	     south-east, the quasar PG 1626+554.}    
    \label{vtp_blend}
\end{figure*}                                                              

\begin{figure}
  \epsfxsize=0.5\textwidth
  \hspace*{0cm} \centerline{\epsffile{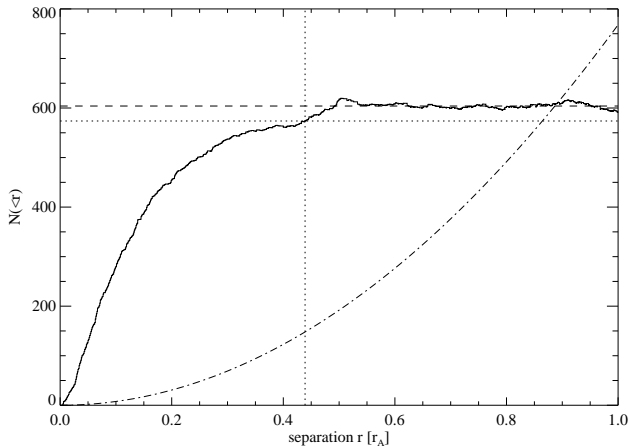}}
  \caption[]{The cumulative number of coincidences in the cross-correlation 
	   between the ACO catalogue and the RASS X-ray source list provided
	   by VTP  as a function of the X-ray to optical source 
           separation. A parabolic background component (the dot-dashed line)
           has been subtracted.}
  \label{vtp_cumnum}
\end{figure}

Subsequently, the merged VTP X-ray source list for all PET fields was
correlated against the ACO catalogue, and, just as before, a 95 per
cent complete sample of coincidences was compiled (see
Fig.~\ref{vtp_cumnum}). The maximal separation between the included
matches is 0.439 Abell radii. This sample now comprises 723 X-ray
sources corresponding to 599 ACO clusters, the higher number being due
to additional VTP detections of nearby clusters missed or misassessed
by the SASS, and also to a number of serendipitous detections of X-ray
faint ACO clusters in the selected PET fields.

\section{Removal of non-cluster sources}
\label{clean_up}

Clearly, this sample cannot be used as it stands, as, statistically,
about 148 (i.e.\ some 21 per cent) of its entries are caused by random
coincidences between the optical ACO cluster positions and non-cluster
X-ray sources.  In fact this figure is still a lower limit since, in
addition to truly random coincidences, we also expect a few percent of
the listed matches to be with non-cluster sources like QSOs or AGN
which are, however, cluster members. In order to identify chance
coincidences as well as cluster-related point sources, we searched the
{\sc Simbad} database in Strasbourg for alternative optical
counterparts for the 723 VTP-detected X-ray sources. For each X-ray
source, we also obtained $12' \times 12'$ optical images from the POSS
and UK Schmidt sky surveys through the extremely useful {\sc SkyView}
facility\footnote{\tt http://skview.gsfc.nasa.gov/skyview.html} and
looked for possible optical identifications. Also, we searched the
ROSAT public archive of pointed PSPC and HRI observations for further
X-ray data for our cluster candidates. The 135 deeper X-ray images
thus obtained plus another seven from non-public pointed observations
awarded to members of the IoA X-ray group were then examined in order
to ensure that the emission detected by VTP in the RASS data
originates indeed from the ICM. The pointed observations were also
used to obtain X-ray fluxes to be used as a control sample for the
VTP-derived RASS fluxes (see Section~\ref{flux_corr}).

As a result of the analysis of all these data,

\begin{itemize} 
 \item 27 of our X-ray sources are identified with stars, 
 \item 30 turn out to be individual galaxies, AGN, or QSOs, and
 \item four are groups or distant clusters of galaxies that are close to 
       but clearly not related to the Abell clusters they have been 
       associated with. 
 \item In 45 cases, the X-ray source cannot be assigned unambiguously to an 
       optical counterpart, but is point-like and not associated with any 
       enhancement in the projected galaxy density, which is why we reject 
       these sources too as unrelated. 
 \item For 18 clusters, the detected emission is heavily contaminated by point 
       sources which may or may not be related to the cluster. We make no 
       attempt at disentangling the clusters' and the contaminants' contribution, 
       and discard these candidates, for all of which the diffuse emission is 
       negligible in comparison to the dominating point source. 
 \item Finally, we eliminate another 41 X-ray sources, most of which are very 
       faint, on the grounds of their being either spurious or not fully 
       contained in the respective PET field. Some of the latter are also (and 
       wholly) contained in a neighbouring field so that in these cases the 
       clusters concerned are actually not lost from the sample.
\end{itemize}

In Table~\ref{noncl_tab} we list the 37 sources that could be bright
enough to get included in our flux limited sample (see
Section~\ref{flx_lim} for a discussion of the XBACs' flux limit of
completeness) but were discarded as non-cluster sources in the above
procedure. Note that our assessment of these sources' brightness is
based on their count rates only, as a rigorous conversion to X-ray 
fluxes requires knowledge of the physical emission process.

\begin{table*}
  \begin{tabular}{lrrll}
        \multicolumn{1}{c}{associated}  & \multicolumn{1}{c}{$\alpha$ (J2000)} & 
                                          \multicolumn{1}{c}{$\delta$ (J2000)} & 
                                                                identification & comments \\
        \multicolumn{1}{c}{ACO cluster} & \multicolumn{1}{c}{[deg]} & 
                                          \multicolumn{1}{c}{[deg]} & & \\ \hline \\      
        & & & & \\
  A2751   & $  4.1560$ & $-31.4213$ & ambiguous      & probable blend of (X-ray faint) cluster and AGN \\
  A27     & $  6.3199$ & $-20.8596$ & probable AGN   & isolated galaxy is most probable id \\
  A2807   & $ 10.1788$ & $-34.6677$ & AGN            & Seyfert~1, z=0.199, Cristiani et al.\ (1995) \\ 
  A2836   & $ 13.2024$ & $-47.6170$ & probable AGN   & on isolated galaxy \\
  A151    & $ 17.2066$ & $-15.6297$ & star           & HD6853, G5, 10.1 mag \\ 
  A2881   & $ 18.1192$ & $-17.0151$ & star           & YZ CET, 12.0 mag dMe star \\
  A2904   & $ 20.5826$ & $-29.2490$ & ambiguous      & probably blend of two stars \\
  A195    & $ 21.5654$ & $ 19.1654$ & star           & HD8723, F0, 5.3 mag \\
  A195    & $ 21.8822$ & $ 19.1838$ & AGN            & MRK 0359, Seyfert~1.5, $v=5007$ km s$^{-1}$ \\
  A195    & $ 21.9834$ & $ 18.9909$ & star           & BD+18 193, F8, 9.2 mag \\
  A271    & $ 28.4562$ & $  1.8797$ & anon star      & 13 mag star most probable id \\
  A347    & $ 36.4376$ & $ 41.9761$ & ambiguous      & NGC911, $v=5604$ km s$^{-1}$ or HD14933, A0, 9.2 mag \\ 
  A3107   & $ 48.7204$ & $-42.6777$ & AGN            & 0313-428, Seyfert~1, $z=$0.126, Monk et al.\ (1988) \\ 
  A484    & $ 63.8387$ & $ -7.6486$ & star           & HD 26965, K1, 4.4 mag \\
  A3367   & $ 87.3101$ & $-24.4209$ & probable AGN   & IRAS 05472-2426 in error box, no published redshift \\
  A3389   & $ 95.7874$ & $-64.6040$ & AGN            & Seyfert~1, IRAS C06229$-$6434, z=0.130 \\
  A3392   & $ 96.7710$ & $-35.4898$ & radio galaxy   & PKS0625-35, $z=0.05459$, core-dominated dominated central galaxy \\
  A3408   & $107.1714$ & $-49.5434$ & AGN            & 4U 0708-49, Seyfert~1, $z=0.0411$ \\
  A689    & $129.3556$ & $ 14.9830$ & prob BLLac     & on flat-spectrum radio source in cluster core, Crawford et al.\ (1995)\\
  A757    & $138.3565$ & $ 47.6871$ & probable AGN   & blend of two AGN candidates \\
  A1030   & $157.7444$ & $ 31.0550$ & AGN            & B2 1028+31.3, $z=0.1782$, at very core of cluster \\
  A1225   & $170.2863$ & $ 53.8547$ & probable AGN   & on isolated blue galaxy \\
  A3494   & $179.2058$ & $-32.2318$ & star           & HD103743, G0, 7.0 mag \\
  A1593   & $190.5454$ & $ 33.2872$ & AGN            & WAS 61, Seyfert~1, $z=0.0439$ \\
  A1599   & $190.7622$ & $ 2.7423 $ & Galaxy         & NGC~4636, $v=1095$ km s$^{-1}$ \\
  A3565   & $203.9744$ & $-34.2951$ & AGN            & MCG-06-30-015, Seyfert~1, $v=2248$ km s$^{-1}$ \\
  A1774   & $205.2705$ & $ 39.9931$ & probable BLLac & B3 1338+402, on flat spectrum radio source in cluster core \\
  A3574   & $207.3311$ & $-30.3166$ & AGN            & IC 4329A, Seyfert~1, $v=4813$ km s$^{-1}$ \\
  A1855   & $211.2495$ & $ 47.1144$ & AGN            & RX~J14050+4707, $z=$0.1510 Bade et al.\ (1995) \\ 
  A2034   & $227.6707$ & $ 33.5958$ & probable AGN   & on isolated galaxy \\
  A2147   & $240.9162$ & $ 15.9002$ & AGN            & B1601+1602, $z=$0.1095 \\
  A3698   & $308.6135$ & $-25.5439$ & star           & PPM 736274, 9.7 mag \\
  A3747   & $317.0595$ & $-43.8522$ & AGN            & Seyfert~1, C08.02, no published redshift, Maza et al.\ (1992) \\
  A2351   & $323.5352$ & $-13.4789$ & star           &  HD 205249, G5,  8.0 mag \\
  A3915   & $341.6812$ & $-52.1109$ & probable AGN   & on isolated galaxy \\
  A2660   & $356.2095$ & $-26.0307$ & probable AGN   & on faint blue, stellar object \\
  A4038   & $357.2267$ & $-28.1217$ & star           & HD 223352, A0, 4.6 mag \\
  \end{tabular}
  \caption[]{Sources with VTP count rates higher than 0.25 s$^{-1}$ that were 
           excluded as being dominated by, if not entirely due to, X-ray 
           emission of non-cluster origin. 
           The quoted positions are the centroids of the VTP detections.}
  \label{noncl_tab}
\end{table*}
 
In the course of the visual inspection of both the X-ray and the
optical images of all our candidate sources we also encountered a few
cases where extended X-ray emission clearly detected in the RASS data
is missing from our tentative list as the respective source is too far
away from the nominal optical cluster position. After careful
examination of the optical and X-ray data we include eight sources
(associated with seven ACO clusters) at X-ray/optical separations of
more than 0.439 Abell radii into our sample. Five of the affected
clusters (A\,548 [twice], A\,1750, A\,1758, A\,1631, and A\,3528) already had
X-ray sources assigned to them in our list and thus join the small
sample of double clusters (for A\,625, A\,901, A\,2197, and A\,2572
both components are already contained in our list) with both X-ray
detections being of comparable brightness in all cases.  For the other
two (A\,2537 and A\,3869) the added detection is the only one, and, in
particular in the case of A\,3869, it is not entirely clear whether
the system detected by ACO in the optical is in fact the same as the
cluster now detected in the X-ray.

\begin{figure}
  \epsfxsize=0.5\textwidth
  \hspace*{0cm} \centerline{\epsffile{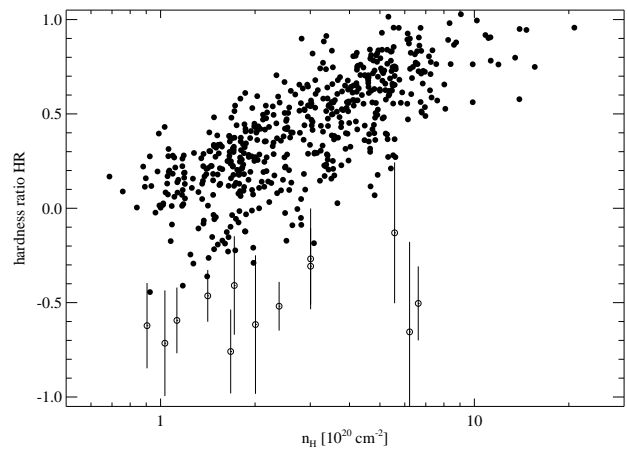}}
  \caption[]{The observed spectral hardness ratio for the 566 X-ray sources
           of the decontaminated sample as a function of the Galactic
 	   column density of neutral Hydrogen.
           Open circles with error bars represent the 13 sources examined
           once more because of their softness.}
  \label{nh_hr}
\end{figure}

So far, we have removed 165 sources from our sample on the grounds of
their combined X-ray/optical appearance. A further criterion that has
not yet been applied is the spectral hardness of the X-ray
emission. We use the SASS definition of a source's hardness ratio HR
\[ 
          {\rm HR = \frac{h-s}{h+s} } 
\]
where $h$ are the photon counts in the hard energy range from 0.5 to 2
keV, and $s$ those in the softer 0.1 to 0.4 keV band. X-ray emission
from a plasma at temperatures of typically $10^{7-8}$ K occurs
predominantly in the hard band; accordingly, the hardness ratios of
RASS-detected clusters of galaxies are found to be positive in more
than 90 per cent of the cases (Ebeling et al.\ 1993).  Although
clusters of galaxies are intrinsically hard X-ray sources, the emitted
spectrum is subject to modification on its way to the observer through
absorption of soft X-rays by neutral Hydrogen in the Galaxy. The
variation of the hardness ratio values observed for the 566 sources
remaining in our sample with the column density of neutral Hydrogen
$n_{\rm H}$ along the line of sight is shown in Fig.~\ref{nh_hr}. (The
$n_{\rm H}$ values are taken from the compilation of Stark et al.\
(1992) for declinations north of $-40$ degrees and from Dickey \&
Lockman (1990) in the remainder of the sky.) As can be seen from
Fig.~\ref{nh_hr}, the observed HR for ICM emission can become small
and even negative where the absorption is extremely
low. Figure~\ref{hr_hist} shows the same distribution of HR values as
a histogram; note the low-HR tail of the distribution. Some sources,
(marked by open circles in Fig.~\ref{nh_hr} and light shading in
Fig.~\ref{hr_hist}) feature HR values much lower than what would be
expected from the corresponding column density, and are thus unlikely
to be due to X-ray emission from galaxy clusters. For most of the 13
soft sources highlighted in Fig.~\ref{nh_hr} the errors in the
hardness ratios are, however, too large to firmly rule out a cluster
origin which is why we scrutinize all 13 once more in both the X-ray
and the optical. As a result,
\begin{itemize} 
  \item we exclude one source on the grounds of its X-ray spectrum
        being too soft to be attributed to ICM emission (6$\sigma$ off the 
 	overall $n_{\rm H}$-HR relation) and
  \item discard another seven point-like sources that can tentatively be 
        identified as stars or AGN.                              
\end{itemize}

\begin{figure}
  \epsfxsize=0.5\textwidth
  \hspace*{0cm} \centerline{\epsffile{xbacs_hr_hist.epsf}}
  \caption[]{The distribution of spectral hardness ratios for the 566 X-ray 
           sources of the decontaminated sample (dark shading).
           The 13 soft sources highlighted in Fig.~\protect\ref{nh_hr}
	   are shown in light shading.}
  \label{hr_hist}
\end{figure}
 
Altogether, we have thus eliminated 173 X-ray sources from our tentative
sample and added in another eight which leaves us with 558 sources.
Fig.~\ref{good_bad_xy} shows the spatial distribution
of the rejected sources with respect to the nominal ACO cluster
positions and compares it to the same distribution for the sources
remaining in the sample. Just as expected for a sample dominated by
random coincidences, the removed sources' distribution is homogeneous,
whereas the distribution of the sources attributed to ICM emission is
found to be strongly peaked around the optical cluster position.

\begin{figure*} 
  \parbox{0.49\textwidth}{
  \epsfxsize=0.49\textwidth
  \epsffile{xbacs_xy_bad.epsf}}
  \parbox{0.49\textwidth}{
  \epsfxsize=0.49\textwidth
  \epsffile{xbacs_xy_good.epsf}}
  \caption[]{The projected spatial distribution of the 173 X-ray sources removed 
           from the sample around the nominal cluster positions (left) and the
           same distribution for the 558 sources accepted as being due to ICM
	   emission (right). The dotted circle marks the maximal X-ray/optical
           separation of 0.439 Abell radii. Data points outside the circle in 
           the 
	   right-hand plot represent components of double clusters or clusters
 	   with unusually large X-ray to optical separations (see text for
	   details).}
  \label{good_bad_xy}
\end{figure*}

The 558 X-ray sources remaining in the cleaned sample are associated
with 532 different ACO clusters (nine of them double), as 48 sources
are classified as multiple detections of the same cluster in the sense
that the extended cluster emission got picked up in more than one
piece. However, contrary to the nine double clusters, the 22 clusters
affected do not consist of clearly distinct subentities, which is why,
for each of these 22, we merge the different sources detected by VTP
into one. Accordingly, the number of X-ray sources drops by 26 to equal
the number of unique ACO clusters, 532.

As can be seen from Figs.~\ref{nh_hr} and \ref{hr_hist}, the sample of
558 still contains a number of X-ray sources that are suspiciously
soft. For many of these, but also for a number of spectrally hard
sources, the X-ray images from the RASS show evidence of contamination
from point sources. We flag a total of 21 clusters for which the level
of contamination might exceed 30 percent of the total flux. For four
of these 21, pointed data are available which will be used in
Section~\ref{cont_test} to quantify the contribution of point sources
to the total source flux.

\section{Getting the fluxes right}

Besides selection effects, the main concern when compiling an X-ray
flux limited sample is the reliability and accuracy of the flux
determination procedure.  This section details how diffuse emission
that has escaped direct detection is taken into account, compares the
resulting RASS-VTP count rates with results obtained from pointed
observations, and finally describes the conversion of count rates into
energy fluxes.

\subsection{Correction for `missing flux'}
\label{flux_corr}

In the presence of background radiation, the emission directly
detectable by any source detection algorithm is always limited to some
fraction of the total flux. The raw VTP count rates for the 523
clusters of our sample (532 if the two components of our nine double
clusters are counted separately) thus have to be corrected for the low
surface brightness emission in the far wings of the source that has
not been detected directly. In doing this, we use King's approximation
to the density profile of an isothermal sphere (King 1962) and assume
a source profile of the form
\begin{equation}
\sigma_{\rm K}(r) = \sigma_0 \left[ 1+(r/r_c)^2 \right]^{-3\beta+1/2} 
          \equiv \sigma_0 \, \tilde{\sigma}_{\rm K}(r)
\label{eq_king}
\end{equation}
(Cavaliere \& Fusco-Femiano 1976) where $\sigma_{\rm K}(r)$ is the
projected surface brightness as a function of radius. Fixing the beta
parameter at a value of 2/3 (Jones \& Forman 1984), we can derive both
the normalization $\sigma_0$ and the core radius $r_c$ from the VTP
source characteristics. Note that it is only in {\em correcting} the
detected count rates that a specific model for the source profile and,
in particular, spherical symmetry is assumed. As this flux correction
step is crucial for any flux-limited cluster sample, the procedure
applied shall be described in some more detail in the following.

The detected and background corrected count rate $s$ can be written as
\begin{equation}
s = s(r_{\rm VTP}) = 2 \pi \, \sigma_0 \int_0^{r_{\rm VTP}} \tilde{\sigma}(r)\,r\,dr.
\label{eq1}
\end{equation}
Here, $r_{\rm VTP}$ is the effective radius of the emission pattern detected
by VTP. The algorithm making no assumptions about the sphericity of sources,
the detected emission can essentially be of arbitrary shape. $r_{\rm VTP}$ 
is then given by
\[
A_{\rm VTP} = \pi \, r_{\rm VTP}^2,
\]
where $A_{\rm VTP}$ is the total area covered by the Voronoi cells of
all photons assigned to any one source by VTP (see Ebeling \&
Wiedenmann 1993 for details of the VTP algorithm). The observed
surface brightness distribution $\tilde{\sigma}(r)$ in Eq.~\ref{eq1}
is the convolution of the source's intrinsic surface brightness
distribution $\tilde{\sigma}_{\rm K}$ with the instrument response:
\[
\tilde{\sigma}(r) = \int_0^{2\pi} \int_0^\infty
                  \tilde{\sigma}_{\rm K}(|\vec{r}-\vec{r}\,'|) 
                  \,{\rm PSF}(r') \, r' \, dr' \, d\phi.
\]
For the RASS, the telescope's point-spread function, PSF($r$), is the
weighted average of the PSFs at all off-axis angles. PSF($r$) has been
computed for several photon energies by Hasinger et al.\ (1994); we
employ a numerical representation of PSF($r$) for $E=1$ keV (De
Grandi, private communication) that was shown to be in excellent
agreement with the source profiles found with RASS-detected AGN
(Molendi 1995).

Besides Eq.~\ref{eq1}, the second crucial equation is the one
specifying how close to the background the surface brightness of the
outermost regions of a VTP detection is:
\begin{equation}
 \sigma_0 \, \tilde{\sigma}(r_{\rm VTP}) = (f-1) \, \sigma_{\rm bkg}.
\label{eq2}
\end{equation}
Both the average surface brightness of the background emission, $\sigma_{\rm 
bkg}$, and $f$, the normalised distance of the lowest surface brightness
region to the background level, are parameters returned by VTP for each source.
From Eqs.~\ref{eq1} and \ref{eq2}, we can thus eliminate $\sigma_0$ in order to
first solve numerically for the core radius $r_c$, and then, with
$\tilde{\sigma}'(r_{\rm VTP})$ known, obtain $\sigma_0$ from Eq.~\ref{eq2}.

Finally, the true total source count rate can be determined from 
\begin{equation}                                                               
s_{\rm true} = 2 \pi \int_0^\infty \sigma_{\rm K}(r) \, r \, dr 
             = \frac{\pi \, \sigma_0 \, r_c^2}{3\,(\beta-1/2)}.
\label{king_true}
\end{equation}                                                               

Note that at no stage of the flux correction procedure the model
profile from Eq.~\ref{eq_king} is actually fitted to the observed
surface brightness distribution. Although we do assume a specific
model, it is only its integral properties that enter, which is why
deviations of the true distribution from the assumed model have much
less effect on the result than they do for the fitting procedures
employed by conventional source detection algorithms.  The lack of any
radial fitting, however, also entails that for any particular cluster
the value for the cluster core radius $r_c$ determined in the flux
correction process can not be expected to be as accurate as what might
be obtained in a detailed imaging analysis of pointed data. Rather, it
should be seen as a statistical parameter that allows the fraction of
the cluster emission that has escaped direct detection to be assessed.

Due to the limited spatial resolution of the PSPC, clusters are
recognized as extended sources only if their projected effective core
radius $r_c$ exceeds some 15 arcsec. For even more compact sources we
compute the flux correction factors from the PSF alone:
\begin{equation}                                                               
 \frac{s(r_{\rm VTP})}{s_{\rm true}} = 
           2 \pi \int_0^{r_{\rm VTP}} {\rm PSF}(r) \, r \, dr.
\label{psf_true}
\end{equation}                                                               

\begin{figure}
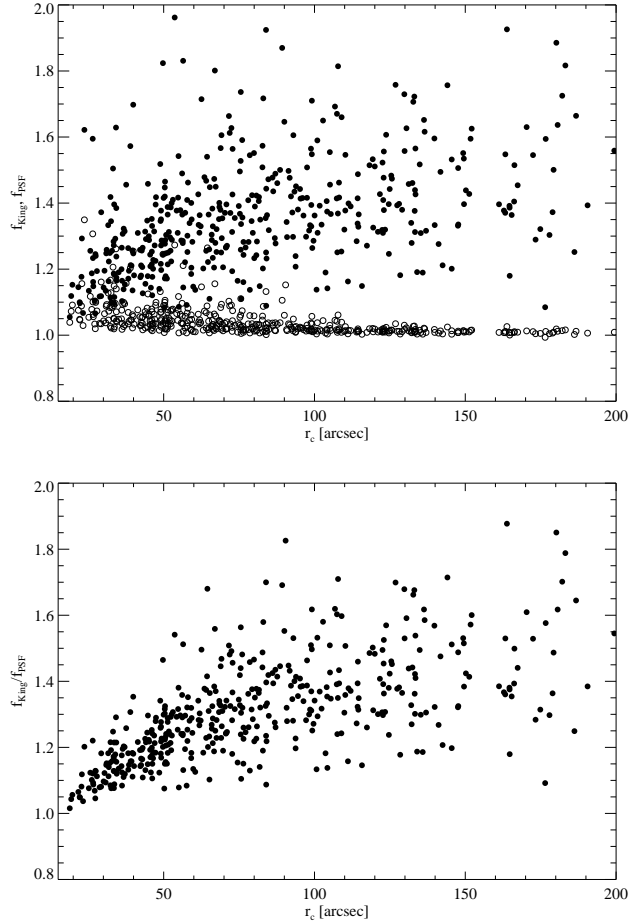

  \epsfxsize=0.5\textwidth
  \hspace*{0cm} \centerline{\epsffile{xbacs_fcorr_rc.epsf}}
  \epsfxsize=0.5\textwidth
  \hspace*{0cm} \centerline{\epsffile{xbacs_fcorr_rat_rc.epsf}}
  \caption[]{The count rate correction factors for our sources as a function
           of the core radius determined for a King profile. The top panel
	   shows the values obtained when either a convolution between a King 
	   law and the instrument PSF (filled circles) or the instrument PSF
	   alone (open circles) is used for a radial surface brightness
	   profile. The bottom plot shows the ratio of the two correction
           factors and illustrates how both solutions converge for
	   point-like sources, i.e., for core radii below some
           20 arcseconds. Eleven sources consisting of less than twenty
           photons have been omitted from these plots.}
  \label{fc_rc}
\end{figure}

Figure~\ref{fc_rc} shows the resulting flux correction factors $s_{\rm
true}/s$ for a King profile convolved with the PSF and the PSF alone
(cf.\ Eqs.~\ref{king_true} and \ref{psf_true}) as a function of the
core radii $r_c$ returned by the above procedure. Note how the flux
correction factors for extended and point-like sources converge at
$r_c \sim 15''$.

\subsection{Correction for flux from point sources and the accuracy of VTP 
            fluxes}
\label{cont_test}

The accuracy of the corrected RASS-VTP count rates can be assessed by
comparing them to the PSPC broad band count rates obtained in pointed
observations (POs) of the same clusters. To this end we retrieve from
the ROSAT public archive the PSPC images and exposure maps for 100
clusters with RASS detections at count rates ranging from 0.1 to 10
counts per second in the PSPC broad band. The median exposure time for
this set of pointings is 10 ks, the extreme values are 1700 s
(comparable to the deepest PET fields from the RASS) for A\,76 and 45
ks for A2218.

Total count rates in the PSPC broad band are obtained independently by
two of us (HE and ACE) by interactively selecting circular cluster and
background regions in the flat-fielded PSPC images. In order to allow
the contribution from point sources to the total cluster count rate to
be quantified, we determine two count rates per cluster, one from
diffuse ICM emission, and a second one from any point sources embedded
therein. The background is taken from regions that appear source-free
in a vignetting corrected image of the respective PSPC field. For all
but the faintest clusters we find the systematic uncertainties of some
5 per cent to dominate over the statistical error of 1 per cent
(median) in the obtained count rates. Note that no VTP derived
parameters are used in this straight-forward, interactive
determination of cluster count rates.

\begin{figure}
  \epsfxsize=0.5\textwidth
  \hspace*{0cm} \centerline{\epsffile{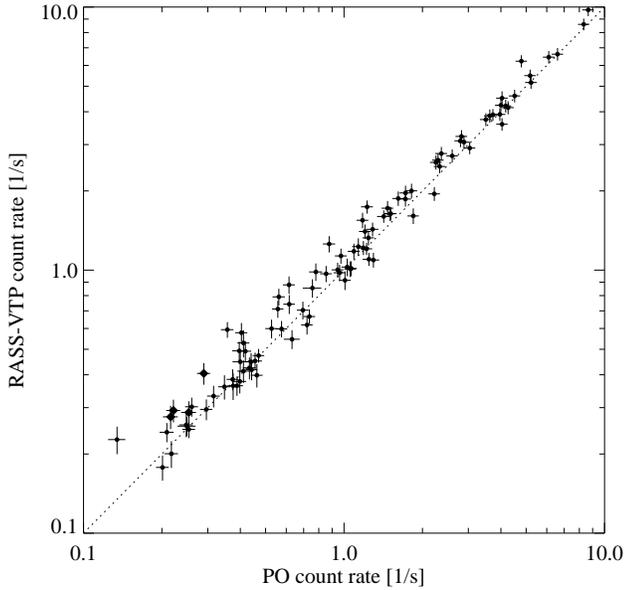}}
  \caption[]{PSPC broad band count rates for 100 ACO galaxy clusters as
	     derived from ROSAT All-Sky Survey data using VTP, compared to 
	     the corresponding values obtained from pointed observations (PO). 
             Note that the contribution from point sources is {\em included}\/ in
             the PO count rates. The error bars allow for a five per cent 
             systematic uncertainty in addition to the Poissonian errors.
             The four clusters shown as filled diamonds have been marked as
	     probably severely contaminated on the grounds of their appearance
	     in the RASS X-ray images.}
  \label{RASS_PO_cr}
\end{figure}

Figure~\ref{RASS_PO_cr} compares the count rates determined from the
pointed observations {\em including point sources}\/ with the
corrected RASS-VTP count rates (cf.~\ref{flux_corr}). The error bars
in Fig.~\ref{RASS_PO_cr} were obtained by adding Poissonian and
systematic errors in quadrature; the latter are assumed to be five per
cent for VTP and PO count rates alike.

Note that, although the overall distribution is in reasonable
agreement with the one-to-one relation expected for perfect agreement
between the two data sets (shown as a dotted line in
Fig.~\ref{RASS_PO_cr}), the count rates obtained from the RASS lie
systematically above the PO values by about nine per cent at all flux
levels. This effect is, however, anticipated. At the sensitivity of
the RASS, X-ray point sources within a cluster, such as individual
galaxies or QSOs, are in general not resolved but blended into the
diffuse X-ray emission from the ICM. As a consequence, $s(r_{\rm
VTP})$ in Eq.~\ref{eq1} is an overestimate of the count rate from ICM
emission within $r_{\rm VTP}$. The overestimation becomes worse when
the correction for not directly detection flux, which assumes a pure
King profile, is applied to this sum of diffuse and point-like
emission. Starting from a contaminated source, the procedure described
in Section~\ref{flux_corr} thus over-corrects the missing ICM flux.

The four clusters shown as solid diamonds in Fig.~\ref{RASS_PO_cr} are
the ones flagged earlier as possibly considerably contaminated by
non-ICM emission {\em on the basis of their appearance in the RASS
alone}\/ (cf.\ Section~\ref{clean_up}). They do, however, not stand
out particularly in Fig.~\ref{RASS_PO_cr} -- in fact, many clusters
that lie even higher above the dotted line (indicating the nominal
count rate) have not been recognized as contaminated in the RASS
data. We therefore resort to correcting for contamination on a
statistical rather than on a cluster-by-cluster basis.

As the point-source contamination affects not only the directly
detected emission but also, and in a complicated way, the flux
correction factor, the resulting systematic overestimation in the
total VTP count rates cannot be corrected for
straightforwardly. However, there is a simple heuristic solution. With
the initially detected VTP count rates being too low and the corrected
ones being too high, we adopt their geometrical mean as the best RASS
determination of the ICM emission from clusters of galaxies.

\begin{figure}
  \epsfxsize=0.5\textwidth
  \hspace*{0cm} \centerline{\epsffile{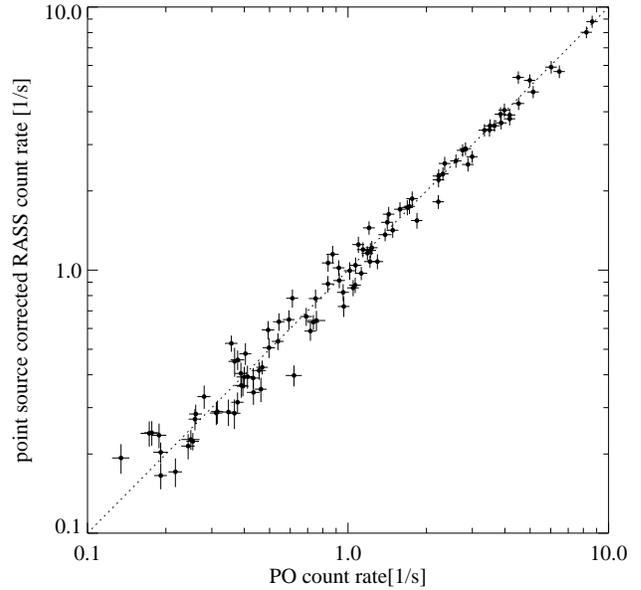}}
  \caption[]{PSPC broad band count rates for 100 ACO galaxy clusters as
	     derived from ROSAT All-Sky Survey data using VTP, compared to 
	     the corresponding values obtained from pointed observations (PO). 
             Note that the contribution from point sources is {\em excluded}\/ in
             the PO count rates. The error bars allow for a five per cent 
             systematic uncertainty in addition to the Poissonian errors.}
  \label{RASS_PO_cr_corr}
\end{figure}

\begin{figure}
  \epsfxsize=0.5\textwidth
  \hspace*{0cm} \centerline{\epsffile{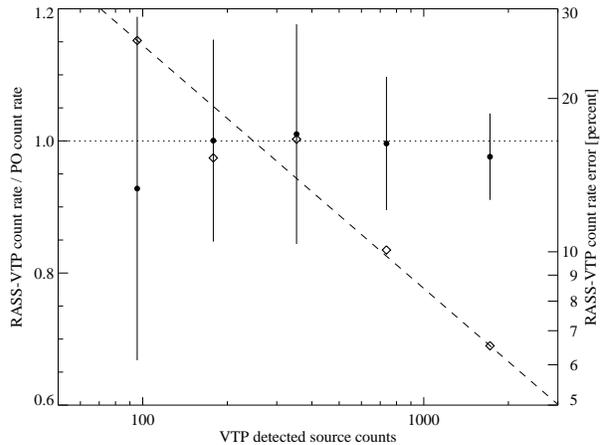}}
  \caption[]{Average ratio of the count rates
	     derived from ROSAT All-Sky Survey data using VTP to 
	     the corresponding values obtained from pointed observations 
             (filled circles with error bars) as a function of the source 
             counts detected by VTP. The sample is the same as in 
             Figs.~\protect\ref{RASS_PO_cr} and \protect\ref{RASS_PO_cr_corr}.
	     Note the excellent agreement between the two sets of cluster
	     count rates at all source count levels. The open diamonds 
	     represent the $1\sigma$ scatter about this one-to-one correlation
	     -- the values refer to the right-hand y axis. The dashed
             line shows the best power-law fit to the $1\sigma$ errors.}
  \label{RASS_PO_cr_err}
\end{figure}

These are the final count rates our X-ray flux limited sample is based
upon.  Figure~\ref{RASS_PO_cr_corr} shows the correlation between the
RASS count rates thus corrected and the corresponding PO values where,
this time, the point source contribution has been removed. The overall
$1\sigma$ scatter of the data points around the one-to-one relation
indicated by the dotted line in Fig.~\ref{RASS_PO_cr_corr} amounts to
15.4 per cent; the mean offset is essentially zero (0.1 per cent).
Note that at no flux level there is a systematic trend for the final
VTP fluxes to over- or underestimate the true fluxes as represented by
the PO values.  This can be seen more clearly from
Fig.~\ref{RASS_PO_cr_err} which shows the ratio of of VTP to PO count
rate as a function of the equivalent VTP source counts; the data
are perfectly consistent with a constant value of unity.
Figure~\ref{RASS_PO_cr_err} also shows how the standard deviation in
the count rate ratio, which represents the fractional $1\sigma$ error
in our final RASS-VTP count rates, decreases with increasing source
photon statistics. From this we find
\begin{equation}
   \frac{\Delta s}{s} = 2.29 \, (s\,t_{\rm exp})^{-0.48},
   \label{vtp_err}
\end{equation}
the dashed line in Fig.~\ref{RASS_PO_cr_err}. The overall, relative
error of our final count rates is thus about twice as high as that
expected from photon statistics alone.

\subsection{Converting PSPC count rates into fluxes}

We convert the corrected count rates to proper energy fluxes using the
XSPEC spectral analysis package and assuming a Raymond-Smith type
spectrum with a global metal abundance of 30 per cent of the solar
value. Values for the Galactic column density of neutral Hydrogen,
$n_{\rm H}$, are taken from the compilation of Stark et al.\ (1992)
for declinations north of $-40$ degrees and from Dickey \& Lockman
(1990) for the rest of the sky. Where available, measured X-ray
temperatures taken from the compilation of David et al.\ (1993) are
used in the conversion (73 clusters); for the remainder the ICM gas
temperature is estimated from the clusters' bolometric X-ray
luminosity. We adopt \[ {\rm k}T = 2.55\,{\rm keV}\, L_{\rm
X,bol,44}^{0.354} \] for the k$T-L_{\rm X}$ relation (D.A.\ White 1996)
where $L_{\rm X,bol,44}$ is the bolometric X-ray
luminosity in units of $10^{44}$ erg s$^{-1}$. Starting from a default
value of 5 keV, k$T$ is then determined from the above relation in an
iteration loop.

\section{Establishing the XBACs' flux limit of completeness}
\label{flx_lim}

\begin{figure}
    \epsfxsize=0.5\textwidth
    \epsffile{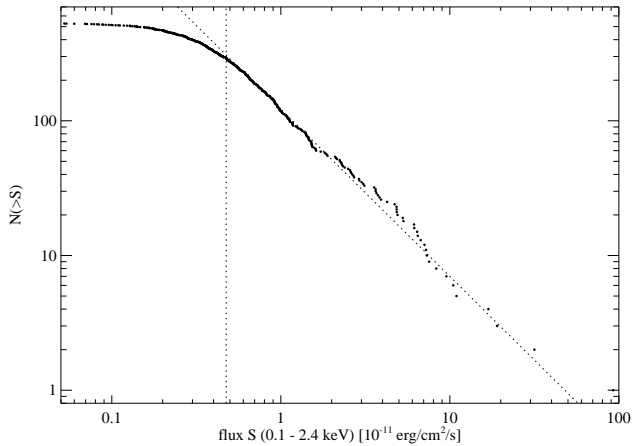}
    \caption[]{The cumulative number count distribution of the 532 X-ray 
           sources constituting our tentative sample as a function of energy 
	   flux in the ROSAT 0.1 to 2.4 keV band. The dotted lines show the
 	   best-fitting power law of slope $-1.25$ and the flux limit 
	   of $5.0 \times 10^{-12}$ erg/cm$^2$/s of a 95 per cent complete 
	   sample.} 
  \label{logn_logs}
\end{figure}

Figure~\ref{logn_logs} shows a simplified $\log N-\log S$ diagram,
namely the cumulative number of clusters as a function of their 0.1 to
2.4 keV flux. The slope of $-1.25 \pm 0.04$ of the best fitting power
law is somewhat steeper than, but not in conflict with, the $-1.02\pm
0.23$ found by Gioia et al.\ (1984) for the EMSS cluster sample.  The
quoted slope is determined in a maximum-likelihood fit of a power law
to the differential, unbinned representation of the data in the range
$S>S_{\rm min}$ where $S_{\rm min}$ is varied from 0.7 to $1.2\times
10^{-11}$ erg/cm$^2$/s. The values of $-1.25 \pm 0.04$ are the mean
and the $1\sigma$ standard deviation of the power law slopes obtained
in this range.

Calling the diagram shown in Fig.~\ref{logn_logs} a `$\log N-\log S$
distribution' is actually not quite correct, as it is not properly
normalized to the sky area covered at any flux value. This is why we
have referred to Fig.~\ref{logn_logs} as a `simplified' $\log N-\log
S$ relation. In order to correctly normalize the distribution, one
would have to know the fraction of sky covered during the RASS at any
limiting flux. The quantity that is relevant in this context is in
fact not the energy flux, but rather the raw count rate at which a
source is initially detected.  As was shown in Section~\ref{rass_db},
sources consisting of at least 13 photons are reliably detected by the
SASS. However, in order to determine accurate fluxes for extended
sources, we re-analysed the RASS data with VTP (cf.\
Section~\ref{ten_sample}), and, in the end, included VTP detections of
sources that had not been detected by the SASS beforehand. These
detections are either of clusters at redshifts below 0.05, where VTP
was run on the RASS images of {\em all} ACO clusters regardless of
whether the SASS had detected them or not, or of clusters that lie
serendipitously in the $2\times 2$ deg$^2$ RASS fields that VTP ran
upon.

It is the combination of two different source detection algorithms and
also that they analysed data sets of, occasionally, very different
exposure times, that make a consistent correction of the $\log N-\log
S$ distribution for sky coverage very difficult. On top of that comes
the incompleteness introduced by the selection effects in the
underlying optical sample; at the lowest X-ray fluxes, this optical
incompleteness is essentially unquantifiable.

Since, however, none of these effects affect the shape of the $\log
N-\log S$ relation at the fluxes used in fitting the power law, we do
not attempt at all to correct for variations in the RASS sky coverage
and the like, and use our simplified $\log N-\log S$ distribution only
to establish the X-ray flux limit of completeness for our sample.

The completeness limit (95 per cent) derived from Fig.~\ref{logn_logs}
is $5.0 \times 10^{-12}$ erg/cm$^2$/s; 277 X-ray sources lie above
this threshold constituting our X-ray flux limited sample of Abell and
ACO clusters of galaxies. Since this flux limit lies significantly
below the flux range over which the power law has been fitted, the
latter is not affected by the increasing incompleteness at fainter
fluxes.

\begin{figure}
    \epsfxsize=0.5\textwidth
    \epsffile{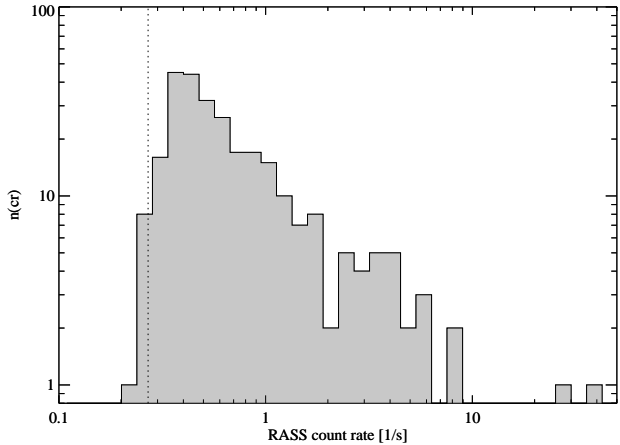}
    \caption[]{The distribution of total VTP count rates for the flux limited
               sample of XBACs sources. The dotted line at a count rate of
  	       0.27 s$^{-1}$ marks the approximate count rate equivalent of
               the flux limit at $5.0 \times 10^{-12}$ erg/cm$^2$/s.}
  \label{xbacs_cr}
\end{figure}

Figure~\ref{xbacs_cr} shows the distribution of total VTP count rates
for these 277 sources. Note that the flux limit of $5.0 \times
10^{-12}$ erg/cm$^2$/s has no direct equivalent in terms of VTP count
rate although it corresponds roughly to 0.27 counts per second (shown
as a dotted line in Fig.~\ref{xbacs_cr}). It is thus crucial that the
flux limit is established properly, i.e., with respect to energy
fluxes rather than detector count rates. The impact of our initial cut
at a SASS count rate of 0.1 s$^{-1}$ on the completeness of our sample
is discussed in Section~\ref{x_incomp}.

The nine clusters classified as double and detected as separate X-ray
sources receive a special treatment. Since we can expect more clusters
in our sample to be physically double systems, which, however, are not
detected as such due to the limited angular resolution of the RASS, we
take this selection effect into account by including into the XBACs
sample those of our double clusters that would meet the flux criterion
if both components were merged, i.e.\ if they were not resolved as
multiple systems by the PSPC. For A\,2572 (Ebeling et al.\ 1995) and
A\,3528 (Raychaudhury et al.\ 1991) this turns out to be unnecessary
as both components are bright enough to be included as clusters in
their own right. For A\,548, A\,901, A\,1631, and A\,1750 only the
flux of the brighter subcluster exceeds $5.0 \times 10^{-12}$
erg/cm$^2$/s; here we include the second component. In the case of
A\,1758 both components fall below the flux limit but are bright
enough to be included when taken together. A\,625 and A\,2197,
finally, are too faint to be accepted even when the flux from both
components is summed.

Having thus included one more cluster, the XBACs sample now comprises
276 clusters seven of which are double. Of the total of 283 clusters
(or subclusters), 277 fall above the flux limit.

\section{The XBACs' Galactic latitude distribution}
\label{xbacs_gb}

\begin{figure} 
  \epsfxsize=0.5\textwidth
  \hspace*{0cm} \centerline{\epsffile{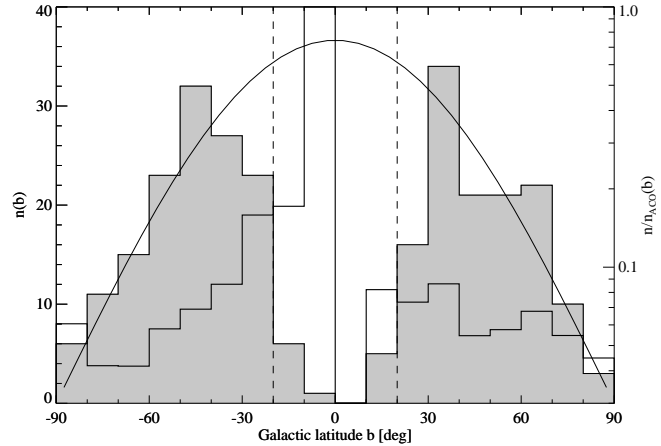} 
                            \hspace*{10mm}}
  \caption[]{The differential Galactic latitude distribution of the 276 clusters
            of our sample of X-ray bright Abell-type clusters (shaded). The
            bold solid line shows to which fraction of all ACO clusters the 
            value for each Galactic latitude interval corresponds; the fine
            solid line indicates the run expected for a uniform distribution
            of clusters on the sky.}
  \label{xbacs_n_gb}
\end{figure}

Figure~\ref{xbacs_n_gb} shows the distribution of the XBACs with
Galactic latitude $b$ (shaded) and also the fraction of all ACO
clusters the XBACs counts in each Galactic latitude bin correspond to
(represented by the bold solid line). The binning ($\Delta b =
10^{\circ}$) was made rather crude deliberately, in order to wash out
all physical large scale structure (the spatial distribution of our
clusters is discussed in Paper III of this series). Note that at $|b|
\ga 30^{\circ}$ the absolute number of clusters in our sample as a
function of $b$ is in good agreement with what is expected for a
uniform distribution of clusters on the sky (shown as the fine solid
line in Fig.~\ref{xbacs_n_gb}). For $|b| \la 30^{\circ}$, however, the
numbers fall significantly short of the predicted values for constant
cluster surface density. Since galaxy clusters do not cluster on
scales of tens of degrees, this effect can not be attributed to
intrinsic variations in the cluster surface density, but rather
reflects ACO's difficulties in reliable detecting galaxy clusters in
the presence of increased Galactic obscuration and stellar density.

As far as the {\em fraction}\/ of all ACO clusters included in our
sample is concerned, we find only a very moderate increase towards the
Galactic plane in the northern Galactic hemisphere. However, the
picture changes dramatically in the southern Galactic hemisphere,
where a steep increase in the fraction of ACO clusters included in our
X-ray flux limited sample is observed within a separation of $\sim
30^{\circ}$ of the Galactic plane. This suggests that, in the south,
ACO's selection was biased in the sense that, with decreasing Galactic
latitude, they included preferentially rich and compact (i.e.,
potentially X-ray luminous) systems in their catalogue.

In order to remain consistent with previous studies (e.g.\ Piccinotti
et al.~1982, Edge et al.~1990, Gioia et al~1990, Henry et al.~1992),
we restrict ourselves to sky areas at high Galactic latitude, i.e.\
$|b| \geq 20^\circ$ (see the dashed lines in Fig.~\ref{xbacs_n_gb}).
Although in regions that far away from the Galactic plane the
incompleteness of the ACO cluster sample due to Galactic obscuration
is small, it is nonetheless still noticeable (cf.\
Fig.~\ref{xbacs_n_gb}). In order to assess the impact of the
incompleteness remaining in the range $20^{\circ} \leq |b| \leq
30^{\circ}$ we flag these clusters in our final list (see
Table~\ref{xbacs_tab}) and test empirically whether the results
presented in the following papers of this series are affected by our
including them or not.

The sky coverage of the high galactic latitude subsample at $|b| \geq
20^\circ$ can be approximated by the geometrical solid angle, i.e.\
$4\pi$ minus the solid angle covered by a 40 degree wide band
representing the Galactic plane. Subtracting another 50 deg$^2$ to
take into account the extragalactic sky obscured by the Magellanic
Clouds yields a solid angle of 8.25 sr.

14 clusters out of the of 532 featured in the simplified $\log N-\log
S$ distribution shown in Fig.~\ref{logn_logs} are at $|b| < 20^\circ$;
the fluxes of 12 of them exceed the flux limit of completeness at $5.0
\times 10^{-12}$ erg/cm$^2$/s. If all 14 are excluded from the sample
before the power law is fitted to the $\log N-\log S$ distribution,
the best fitting slope comes out slightly steeper at $-1.28 \pm 0.04$
which is mainly due to the well-known fact that 3 of the 6 X-ray
brightest ACO clusters (A\,426, A\,3627, and A\,2319) are at low
Galactic latitude. When the low Galactic latitude clusters are
excluded, the flux limit of completeness increases somewhat from 5.0
to $5.3 \times 10^{-12}$ erg/cm$^2$/s which would exclude 16 clusters
more. It thus makes a difference for the XBACs sample 
whether the flux limit or the galactic latitude cut is
applied first. We adopt the former procedure and are left with 271
X-ray sources corresponding to 264 Abell-type clusters.

If finally only clusters at $|b| \geq 30^\circ$ are considered, the
power law of the $\log N-\log S$ distribution of Fig.~\ref{logn_logs}
continues to steepen; the best-fitting slope is $-1.46 \pm 0.03$. Much
of this steepening is, however, again mainly due to a few very X-ray
bright (i.e.\ local) clusters lacking from the high Galactic latitude
sample. No more than about three additional clusters at fluxes $\ga
10^{-10}$ erg cm$^2$ s$^{-1}$ are required to make the best fitting
slope power law consistent with the value of $-1.24$ found for the
$\log N-\log S$ distribution of the overall sample. The apparent
steepening of the $\log N-\log S$ power law can thus be explained
entirely by, e.g., a statistically insignificant dearth of optically
rich enough clusters within a redshift of $\sim$ 0.02.
                                                                   
\section{The XBACs' redshift distribution}
\label{xbacs_z}

As mentioned earlier, the ACO catalogue is known to suffer from
various selection biases inherent to the procedures employed to find
clusters in the optical. In particular, the probability of a cluster
being included in the ACO catalogue is a strong function of redshift:
The more distant an Abell cluster, the richer and more compact it has
to be to be recognized as such on the optical plates. Although the ACO
catalogue was designed to be volume complete out to a limiting
redshift of $z=0.2$, this goal has been achieved only for the richer
systems. Clusters in richness class 0 and 1 start to become
undersampled at redshifts of about 0.13 (e.g.\ Huchra et al.\ 1990,
Scaramella et al.\ 1991, Ebeling et al.\ 1993), i.e.\ at redshifts
higher than that the ACO catalogue is volume complete only for the
richest systems. In an X-ray flux limited sample, however, the
correlation between optical richness and X-ray luminosity (Briel \&
Henry 1993; see also Paper II of this series) tends to counteract this
selection effect.

\begin{figure} 
  \epsfxsize=0.5\textwidth
  \hspace{0cm} \centerline{\epsffile{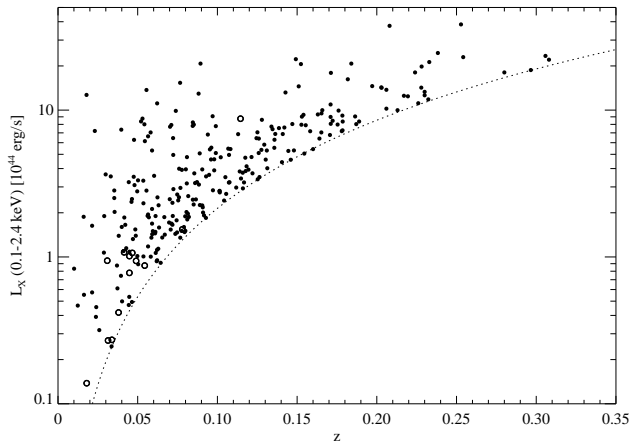}}
  \caption[]{The X-ray luminosities of the 276 clusters of our flux limited 
	    sample of X-ray bright Abell-type clusters (no Galactic latitude
            cut) as a function of redshift. The dotted line illustrates the 
            cutoff introduced by the X-ray flux limit at $5.0 \times 10^{-12}$
            erg/cm$^2$/s. The thirteen VTP detected clusters with original 
            SASS count rates of less than 0.1 s$^{-1}$ are plotted as open 
            circles.}
  \label{xbacs_lum_z}
\end{figure}

Figure~\ref{xbacs_lum_z} illustrates this by showing the XBACs' X-ray
luminosity as a function of redshift: At low and intermediate
redshifts where the ACO catalogue is indeed volume complete, all
luminosity and richness classes get included. With increasing
redshift, however, the X-ray flux limit selects increasingly luminous
(and intrinsically rich) systems disregarding the poorer systems. It
is only for the latter, however, that the ACO catalogue becomes
seriously incomplete at redshifts around 0.13. Consequently, we can
expect our sample to remain unaffected by the incompleteness of the
ACO catalogue out to redshifts considerably beyond the quoted value of
0.13.

\begin{figure} 
  \epsfxsize=0.5\textwidth
  \hspace*{0cm} \centerline{\epsffile{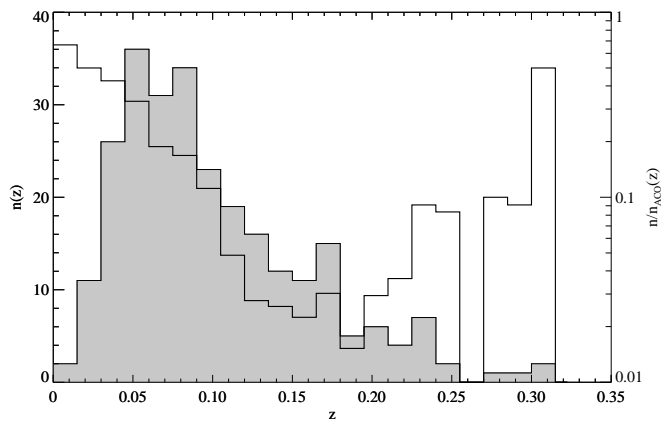} 
                            \hspace*{10mm}}
  \caption[]{The differential redshift distribution of the 264 clusters
            of our sample of X-ray bright Abell-type clusters at $|b| 
            \geq 20^{\circ}$ (shaded). The
            bold line shows to which fraction of all ACO clusters the value
            for each redshift interval corresponds. Both measured and estimated
	    redshifts have been used.}                    
  \label{xbacs_n_z}
\end{figure}

Plotting the redshift histogram for our flux limited sample
(Fig.~\ref{xbacs_n_z}) we find the absolute number of clusters per
redshift bin to drop smoothly with $z$. The {\em fraction} of ACO
clusters included in our sample in each redshift bin, however, starts
to rise at about $z=0.2$, indicative of a severe volume incompleteness
of the underlying optical catalogue {\em if all richness classes
(including 0) are considered}. This rise in the fraction of X-ray
detected clusters occurs at much lower redshifts if no flux limit is
imposed (Ebeling et al.\ 1993).

In order to minimize any remaining impact of the overall
incompleteness of the underlying optical catalogue onto our X-ray flux
limited sample we restrict its redshift range to the nominal
completeness limit of the ACO catalogue at $z=0.2$. The resulting,
`statistical' XBACs sample consists of 242 clusters at $|b| \geq
20^{\circ}$, six of which are marked as double (the seventh double
cluster, A\,1758, at a redshift of 0.28 is now excluded). The issue of
possible incompleteness introduced into the XBACs sample by the volume
incompleteness of the ACO catalogue is discussed in more detail in
Section~\ref{xbacs_comp}.

\section{The completeness of our sample}
\label{xbacs_comp}

Several effects limit the completeness of our flux limited sample most
of which are intrinsic to the way the XBACs have been selected in the
X-ray. Besides that, there is, however, also the important question as
to which extent the XBACs sample is affected by the volume
incompleteness of the optical ACO cluster catalogue.

\subsection{Incompleteness introduced in the X-ray selection procedure}
\label{x_incomp}

As far as the compilation of our sample in the X-ray is concerned,
there are four sources of incompleteness.

First of all, there is the dependence of the RASS sky coverage on the
exposure time mentioned in Section \ref{ten_sample}. At the chosen
minimal exposure time of 130 s, the sky coverage is 95 per cent.

Secondly, clusters are missing from our sample due to the systematic
positional differences between the centroids of the clusters' galaxy
distribution which determine the optical cluster positions and the
centres of the clusters' gravitational potential wells which is what
the X-ray positions correspond to.  These differences are reflected in
the fact that a small fraction of true coincidences between ACO
cluster positions and RASS X-ray sources are found at respective
separations greater than 0.439 $r_{\rm A}$ (see
Fig.~\ref{vtp_cumnum}). The completeness of our sample with respect to
this effect is limited by the finite cut-off in the allowed
X-ray/optical separation and amounts, again, to 95 per cent.

Thirdly, clusters may be missing from our sample due to the
imperfections of the SASS source detection algorithms mentioned
earlier. At redshifts smaller than 0.05 this is not an issue since the
PET data around {\em all}\/ ACO clusters in that redshift range have
been reprocessed by VTP. At higher redshifts, however, clusters that
should be in our flux limited sample may be missing because the SASS
returned count rates that were too low by more than a factor of $\sim
3$. This can be said because the flux limit of our sample of $5.0
\times 10^{-12}$ erg/cm$^2$/s corresponds approximately to a VTP count
rate of 0.27 s$^{-1}$ (cf.\ Fig.~\ref{xbacs_cr}), i.e., a value about
three times higher than the threshold of 0.1 s$^{-1}$ applied to the
initial RASS source list delivered by the SASS. The incompleteness
introduced by this effect can be estimated from the distribution of
SASS and VTP count rates for clusters detected by both algorithms.

\begin{figure} 
   \epsfxsize=0.5\textwidth 
   \hspace{0cm}   \centerline{\epsffile{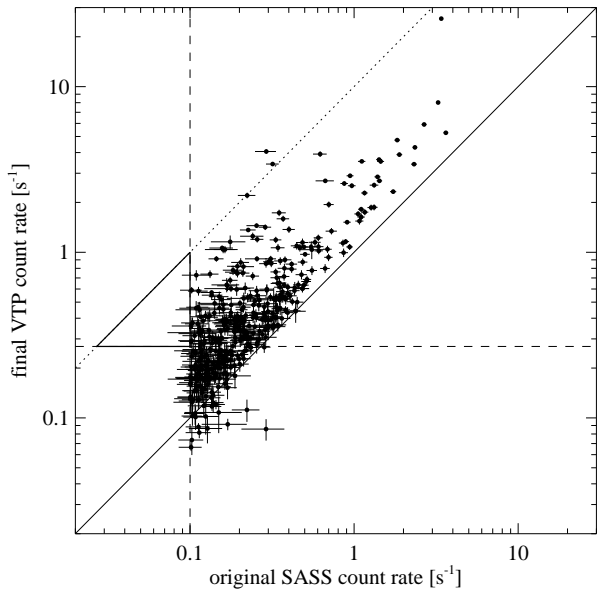}} 
   \caption[]{The relation between original SASS and final VTP count rates for 
            425 ACO clusters detected by both algorithms; the error bars
            represent photon statistics only. The dashed lines mark
            the initial SASS count rate threshold at 0.1 s$^{-1}$ and the
            approximate VTP count rate equivalent of the XBACs flux limit at
            0.27 s$^{-1}$. The bold triangle indicates the region where clusters
            missing from the XBACs sample are expected to lie.}
   \label{vtp_sass_cr}
\end{figure}

Figure~\ref{vtp_sass_cr} shows this distribution; the respective count rate
limits are shown as dashed lines. The triangle on the left of the plot
delineates the section of parameter space  where clusters missing from our
sample reside. Extrapolating from the density of data points to the upper right
of this area, we estimate the number of clusters in this region to be less than
31. Since 13 of those have entered our sample as VTP detections either at low
$z$ (ten clusters) or serendipitously at redshifts above 0.05 (three), the
total number of potentially missing cluster comes down to less than 18, i.e.,
the completeness of our sample with respect to this effect is at least 93 per
cent. 

\begin{figure} 
   \epsfxsize=0.5\textwidth 
   \hspace{0cm}   \centerline{\epsffile{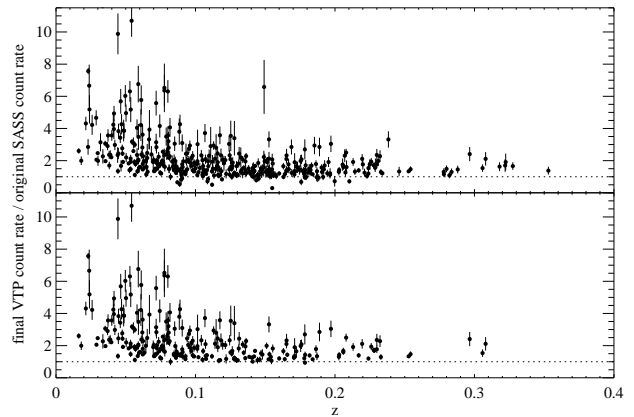}} 
   \caption[]{The ratio of final VTP count rate to original SASS count rate
            as a function of $z$ for the 425 ACO clusters detected by both 
            algorithms (top panel). The lower panel shows the same distribution
            for the clusters that got included in the final XBACs list before
            the redshift cutoff is introduced.
            In both panels the error bars represent photon statistics only.}
   \label{vtp_sass_cr_z}
\end{figure}

That this value is indeed a lower limit can be seen from an
alternative approach. For any cluster featuring a SASS count rate of
less than 0.1 s$^{-1}$, the final VTP count rate would have to be
higher by at least a factor of 2.7 for it to be included in the XBACs
sample. How likely is this to happen?  Figure~\ref{vtp_sass_cr_z}
shows this count rate ratio as a function of redshift for the clusters
that have both SASS and VTP count rates (top panel) and for those that
make it into the XBACs sample before the redshift cutoff at $z=0.2$ is
applied (bottom panel). Of the 425 ACO cluster whose count
rates are shown in Figs.~\ref{vtp_sass_cr} and \ref{vtp_sass_cr_z}
(top panel) no more than 76 fulfill this criterion; 26 of them (i.e.\
about one third) have redshifts less than 0.05. Since we know that at
redshifts below 0.05 {\em the XBACs sample is complete by design}, and
that the number of clusters with $z \leq 0.05$ that made it into the
sample coming from an initial SASS count rate of less than 0.1
s$^{-1}$ is ten, we can extrapolate to a total number of 29 clusters
that should get included despite their SASS count rates falling short
of the threshold value. 13 of these 29 have been detected by VTP, so
that no more than about 16 can be missing. Consequently, we arrive at
an incompleteness of at most six per cent due to our initial cut in
the SASS count rate, consistent with the above estimate. From the
positions of the 13 additional VTP detections in the $L_{\rm X}-z$
distributions of Fig.~\ref{xbacs_lum_z} we conclude that the vast
majority of the missing clusters should feature low X-ray
luminosities, and in any case lie close to the XBACs' flux limit.

According to the distribution shown in Fig.~\ref{vtp_sass_cr_z}, 11 of
the 16 clusters that could be missing can be expected to lie in the
redshift range from 0.05 to 0.1; the ramaining five are anticipated to
be at $0.1 < z \leq 0.2$. The only safe way to find all missing
clusters would be to follow the approach taken for the nearby systems
at $z \leq 0.05$, i.e., re-process the PET fields around {\em all}\/
ACO fields with VTP. This, however, is clearly prohibitive in view of
the massive overhead involved: just within $z=0.1$, 446 {\em
additional}\/ PET fields would have to be re-processed and checked for
non-cluster sources to detect the ten clusters possibly missing in
that redshift range.

Fourthly and finally, our sample is incomplete due to the X-ray flux
limit which was set at `only' 95 per cent completeness.

Since all these effects are most probably statistically independent,
we can simply multiply the corresponding percentages to find the
overall completeness of our sample of X-ray bright Abell-type clusters
of galaxies (XBACs) to be about 80 per cent. For the reasons detailed
above, we believe that this figure comes very close to the maximum
value achievable for such a sample.

\subsection{The impact of optical selection effects}

As mentioned in Section~\ref{xbacs_z}, the X-ray flux limit of the
XBACs sample ensures that, with increasing redshift, more and more
luminous clusters are selected (Fig.~\ref{xbacs_lum_z}). As a
consequence, our sample is, to first order, unaffected by the growing
volume incompleteness in the lower Abell richness classes. At any
redshift, the clusters missing from the ACO catalogue would not have
been X-ray luminous enough anyway to get included in the XBACs sample.

\begin{figure} 
  \epsfxsize=0.5\textwidth
  \hspace{0cm} \centerline{\epsffile{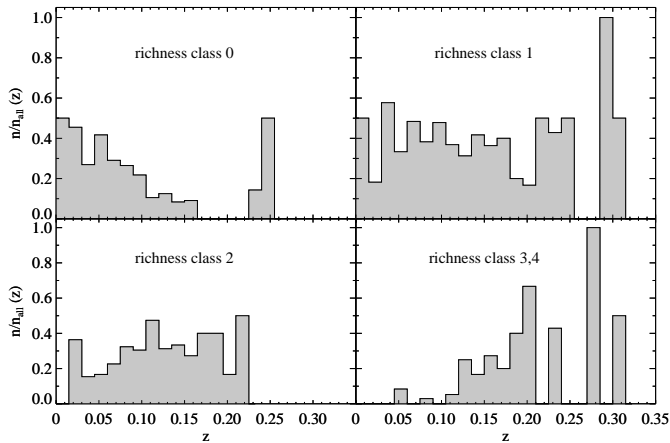}}
  \caption[]{The fraction of XBACs (no redshift cut applied) in 
           different Abell richness
           classes as a function of redshift.}
  \label{xbacs_rc_z}
\end{figure}

Figure~\ref{xbacs_rc_z} illustrates once more how the $L_X-n_{\rm
gal}$ correlation in combination with the X-ray flux limit ensures
that the incompleteness in poor clusters at high redshift does not
affect our sample.  (We drop the redshift limit at $z=0.2$ for the
purpose of this exercise.)  Note how the higher richness classes
dominate more and more as redshift increases; the optically poorer
clusters which tend to be under-represented in the ACO catalogue at
its upper redshift end do not enter the XBACs sample.

It is also noteworthy, however, that of the 13 XBACs at
redshifts higher than 0.225, eight have been classified by ACO as
richness 0 or 1 systems. Given that the X-ray luminosity of all of
them exceeds $10^{45}$ erg s$^{-1}$, it is clear that this
classification must grossly underestimate these systems' true
richness. On the other extreme, we expect the richness of some of the
richness class 2 and 3 systems at redshifts below 0.05 to be
overestimated. For two ACO clusters in that richness and redshift
range this is indeed confirmed by spectroscopic follow-up observations
which yield velocity dispersions of the cluster galaxies that are more
typical of richness class 1 systems (Ebeling \& Mendes de Oliveira, in
preparation).

We conclude that at redshifts below about 0.05 our sample is probably
even more heavily dominated by optically poor clusters than is
apparent from Fig.~\ref{xbacs_rc_z}. At redshifts beyond $\sim 0.15$,
however, where ACO's optical cluster richness starts to be
systematically underestimated, the majority of the XBACs are in fact
of richness class 2 or greater.

\begin{figure} 
  \epsfxsize=0.5\textwidth
  \hspace{0cm} \centerline{\epsffile{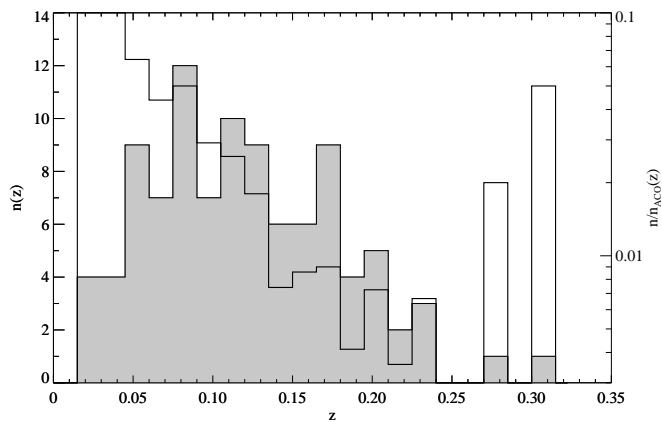}
                            \hspace*{10mm}}
  \caption[]{The differential redshift distribution (before the redshift cut
            at 0.2 is applied) of the 99 XBACs at high Galactic
            latitude and with Abell richness greater than or equal to 2. The
            bold line shows to which fraction of all ACO clusters in the same
	    richness classes the value for each redshift interval corresponds.
	    Both measured and estimated redshifts have been used.}          
  \label{xbacs_n_z_rich}
\end{figure}

Figure~\ref{xbacs_n_z_rich} shows again the redshift distribution of the XBACs
sample (cf.~Fig.~\ref{xbacs_n_z}); this time, however, only the 99 clusters
with $n_{\rm gal} \geq 80$ (i.e., richness classes 2 and greater) are
considered which dominate the sample at $z \ga 0.15$. Note that for these rich
and X-ray luminous systems the ACO catalogue is actually complete out to
redshifts of about 0.23. 

It is thus the X-ray flux limit that makes all the difference when we ask what
impact the optical selection effects have on an X-ray selected sample: The
lower the X-ray flux limit, the more low richness clusters get included and the
higher is the risk of becoming affected by the volume incompleteness of the
optical catalogues. Consequently, the impact of optical selection effects is
strongest for deep X-ray samples  with no single, overall flux limit like the 
ones used in the most recent studies on the X-ray properties of Abell clusters
(Briel \& Henry 1993, Burg et al.\ 1994).

As far as the XBACs are concerned, we cannot exclude rigorously that some
incompleteness due to the imperfections of the underlying optical catalogue
remains. However, based on the above we believe it safe to assume that any
remaining incompleteness will be at redshifts below $\sim 0.1$. At such low to
intermediate redshifts and X-ray luminosities around and below $10^{44}$ erg
s$^{-1}$ we may be missing a few clusters that should be in the ACO catalogue 
on the ground of their optical richness and are, at the same time, sufficiently
X-ray luminous to be included in the XBACs sample. The Virgo cluster, 
explicitly excluded from Abell's catalogue because of its large angular extent,
is one such case (see B\"ohringer et al.\ 1995 for a discussion of the RASS 
data).
 
\section{The XBACS X-ray source extent distribution}

In principle, the ideal X-ray selection criterion for the compilation
of a cluster sample is the X-ray source extent. Even at a redshift of
$z=0.3$, a metric cluster core radius of 250 kpc still corresponds to
45 arcsec on an angular scale. Since this is well above the spatial
resolution of some 20 arcsec attained typically in the RASS, the X-ray
emission from clusters of galaxies is expected to be recognized as
extended over the whole redshift range of the XBACs sample. However,
in practice the efficiency of a source selection by X-ray extent
depends heavily on the sources' brightness and also on the detection
algorithm used.

Although the source extent is not utilized for the compilation of the
XBACs sample, it is relevant for future, X-ray selected samples such
as the ROSAT Brightest Cluster Sample (BCS, Ebeling et al.\ 1996), the
compilation of extended RASS sources discussed by Giacconi \& Burg
(1993), and the ESO Key Programme on ROSAT clusters in the southern
hemisphere (Guzzo et al. 1995).

\begin{figure*}
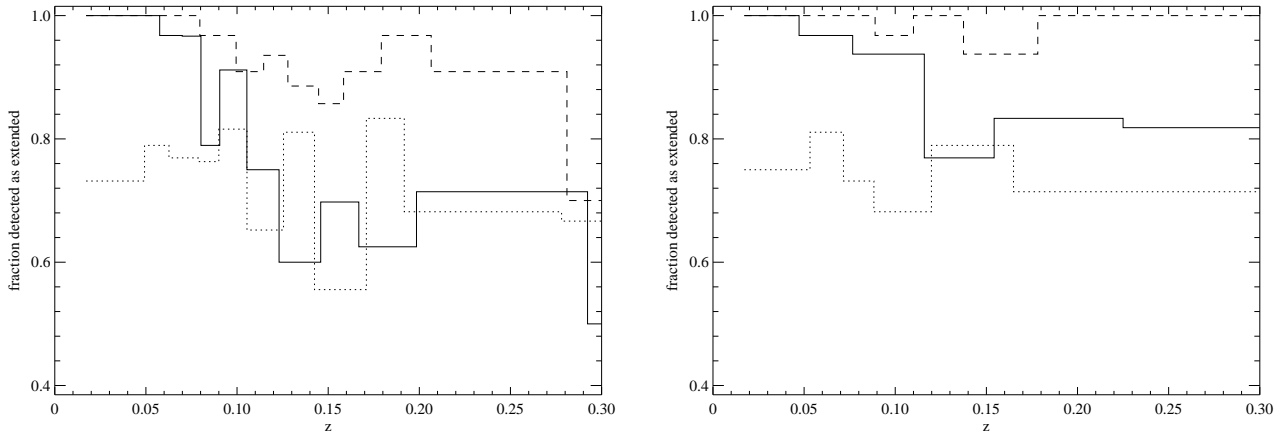
 
   \parbox{0.5\textwidth}{
   \epsfxsize=0.49\textwidth 
   \epsffile{vtp_sass_ext.epsf}} 
   \parbox{0.5\textwidth}{
   \epsfxsize=0.49\textwidth 
   \epsffile{xbacs_vtp_sass_ext.epsf}} 
   \caption[]{The fraction of ACO clusters detected by both SASS and VTP
            that feature an X-ray source extent of more than 35 arcsec as
            a function of cluster redshift. In the left hand plot all 425
            clusters shown in Fig~\ref{vtp_sass_cr} are included.
            The right hand graph uses only
            the 245 ACO clusters that remain when the XBACs flux limit is
            applied. In both plots, each redshift interval (except the very 
            last one) contains 30 clusters; the VTP curves are shown as 
	    solid lines, the SASS data as the dotted lines, and the dashed
            lines represent the fraction of clusters classified as extended
            by either or both of the two algorithms.}
   \label{xbacs_ext}
\end{figure*}

SASS determines source fluxes by fitting a radial Gaussian profile to
the observed spatial photon distribution in a maximum-likelihood
procedure. The extent parameter returned for each source is
essentially the width of this Gaussian that is in excess of the width
of the instrumental point spread function (Cruddace et al.\ 1991). At
a threshold value of 35 arcsec for the Gaussian source extent, half of
all ACO clusters but less than 7 per cent of the non-cluster
sources detected by the SASS in the RASS are classified as extended
(Ebeling et al.\ 1993). A lower extent threshold results in an only
marginally higher fraction of extended clusters while increasing
dramatically the number of point sources that are erroneously
classified as extended.  If only sources with SASS count rates higher
than 0.1 s$^{-1}$ are considered (namely the 425 of
Fig.~\ref{vtp_sass_cr}), the fraction of ACO clusters detected as
extended by the SASS rises to some 73 per cent.  Within $z=0.3$, this
percentage is essentially independent of cluster redshift (dotted
curves in Fig.~\ref{xbacs_ext}).  As the SASS count rate correlates only poorly
with the true cluster count rate (cf.\ Fig.~\ref{vtp_sass_cr}), the
fraction of clusters detected as extended by the SASS does not
increase any further when the XBACs flux limit is applied.  Lowering
the extent threshold does not change much either: even for a threshold
value as low as 10 arcsec the overall figure remains essentially
the same (77 per cent). 

Note that these values are actually upper limits since they do not
take into account the 13 XBACs that were missing altogether
from the SASS source list that we started from.  For the XBACs
within $z=0.05$ these SASS non-detections introduce an {\em
additional}\/ incompleteness of 20 per cent. The reduced detection 
efficiency of the SASS at low redshifts is likely to result in an even 
more severe incompleteness when clusters poorer than those compiled 
by ACO are considered.  

VTP determines a count rate correction factor by comparing the
directly detected count rate with that expected from an idealized
cluster (see Section~\ref{flux_corr}). Besides the actual correction
factor, this procedure also returns the angular equivalent of the
cluster X-ray core radius. It turns out (Ebeling et al., in
preparation) that for this VTP source extent the transition between
clusters and point-like X-ray sources occurs at the same numerical
value (35 arcsec) as before for the SASS. The fraction of VTP extended
sources in the subset of 425 ACO clusters detected by both SASS and
VTP is 79 per cent. It depends on the cluster redshift in the sense
that nearby systems are more likely to be recognized as extended than
distant ones (solid curves in Fig.~\ref{xbacs_ext}). When the XBACs
flux limit is applied, the fraction of clusters classified as extended
by VTP rises to some 80 per cent at $z=0.3$; within a redshift of 0.05
100 per cent are detected as extended. The overall fraction of XBACs
clusters in the flux limited sample that are detected as extended by
VTP exceeds 90 per cent.

At redshifts higher than, say, $z=0.15$, SASS and VTP are similarly
efficient in detecting clusters as extended sources. However, the
systems they classify as extended are not the same. The dashed lines
in Fig.~\ref{xbacs_ext} illustrates the complementarity of the two
algorithms by showing the fraction of clusters recognized as extended
by either SASS or VTP (or both). Only four out of the 245 clusters that the
right-hand plot in Fig.~\ref{xbacs_ext} is based upon are not extended
if the SASS and VTP extended samples are combined.
 
To summarize, we find the X-ray source extent to be a useful parameter
for the selection of clusters of galaxies from RASS data. For a
minimum value of 35 arcsec for the extent parameter and for sources
well above the detection limit, the efficiency is some 70 per cent for
both SASS and VTP {\em for clusters at $z \ga 0.12$}.  At lower
redshifts, any compilation based on SASS source extent alone will be
missing more than one in five ACO clusters (and most probably a much
higher fraction of optically poorer systems). For really nearby
clusters ($z\la 0.05$) the relative insensitivity of the SASS
detection algorithms to low surface brightness emission may result in
a deficiency of more than 40 per cent in the number of the detectable
extended cluster sources. This incompleteness can not be overcome by
the introduction of an X-ray flux limit. If the VTP extent parameter
is used, the incompleteness of a cluster sample selected by X-ray
extent can be reduced to less than 9 per cent at $z\la 0.12$; for the
X-ray flux limited XBACs sample, the overall incompleteness within
$z=0.3$ would be less than 10 per cent if only VTP extended sources
were considered.

If, finally, the extent values from both algorithms are combined,
essentially all XBACs (98 per cent) are classified as extended.

\section{XBACs -- the sample}

Table~\ref{xbacs_tab} lists our statistical sample of the 242 X-ray brightest
Abell-type clusters (XBACs) with Galactic latitudes $|b| \geq 20^{\circ}$,
within a redshift of 0.2, and with 0.1 to 2.4 keV fluxes greater than $5.0
\times 10^{-12}$ erg cm$^{-2}$ s$^{-1}$. Six of these 242 systems consist of two
separate subclusters. 92 per cent of the clusters in this sample have
measured redshifts. We also include the 12 clusters at low Galactic
latitude (marked with a $\bullet$ symbol in the first column) and the 24
clusters at redshifts greater than 0.2 (marked with a $\star$ symbol) that 
meet the flux criterion.

\begin{table}
 \caption[]{not available for preprinting -- sorry!}
 \label{xbacs_tab}
\end{table}

In detail the contents of Table~\ref{xbacs_tab} are
\begin{list}{}{\labelwidth17mm \leftmargin17mm}
  \item[column\,\,\, 1:] redshift, low Galactic latitude, and contamination flag. 
		   Clusters at 
		   $|b| < 20^{\circ}$ ($20^{\circ} < |b| < 30^{\circ}$) are 
		   marked $\bullet$ ($\circ$); $\star$ indicates that the
		   cluster is at $z>0.2$ and thus not a member of our
		   statistical sample; `c' means a significant fraction of
 		   the quoted flux may come from embedded point sources.
  \item[column\,\,\, 2:] ACO name. Where clusters appear to consist of two 
		   components, two entries (`a' and `b') are listed.
  \item[column\,\,\, 3:] Right Ascension (J2000) of the X-ray position as
                   determined by VTP.
  \item[column\,\,\, 4:] Declination (J2000) of the X-ray position as determined
                   by VTP.
  \item[column\,\,\, 5:] Column density of Galactic Hydrogen from Stark et al. (1992)
                         for $\delta \geq -40^{\circ}$ and Dickey \& Lockman (1990) for 
                         the remainder of the sky.
  \item[column\,\,\, 6:] RASS exposure time (accumulated).
  \item[column\,\,\, 7:] PSPC count rate in PHA channels 11 to 235 originally
                   detected by VTP.
  \item[column\,\,\, 8:] The equivalent radius $\sqrt{A_{\rm VTP}/\pi}$ of
                   the source detected by VTP. 
  \item[column\,\,\, 9:] Final PSPC count rate in PHA channels 11 to 235
                   based on the original VTP count rate. Statistical corrections 
                   for low-surface
                   brightness emission that has not been detected directly and
                   for contamination from point sources have been applied.
  \item[column 10:] Error in the final PSPC count rate according to Eq.~\ref{vtp_err}.
                    The fractional uncertainty in the energy flux (column 13)
                    and the X-ray luminosity (column 14) can be assumed to
    		    be the same as the fractional count rate error.
  \item[column 11:] ICM gas temperature used in the conversion from count
 		   rates to energy fluxes. `e' indicates the temperature has
		   been estimated from the $L_{\rm X}-{\rm k}T$ relation.
  \item[column 12:] Redshift. `e' indicates the redshift has been estimated
 		   from the magnitude of the tenth-ranked cluster galaxy.
  \item[column 13:] Unabsorbed X-ray energy flux in the 0.1 to 2.4 keV band.
  \item[column 14:] Intrinsic X-ray luminosity in the 0.1 to 2.4 keV band
			 (rest frame).
  \item[column 15:] Reference for the redshift in column 8.
\end{list}

To facilitate the comparison with previous flux limited samples, we list 
in Table~\ref{2to10_tab} the $2-10$ keV fluxes of the XBACs. The conversion 
from the PSPC detection band was performed using a Raymond-Smith type spectrum
with constant metallicity of 30 per cent of the solar value and values for the
neutral Hydrogen column densities and ICM gas temperatures as given in 
Table~\ref{xbacs_tab}.

\begin{table*}
 \caption[]{not available for preprinting -- sorry!}
 \label{2to10_tab}
\end{table*}

\begin{figure} 
  \epsfxsize=0.5\textwidth
  \hspace{0cm} \centerline{\epsffile{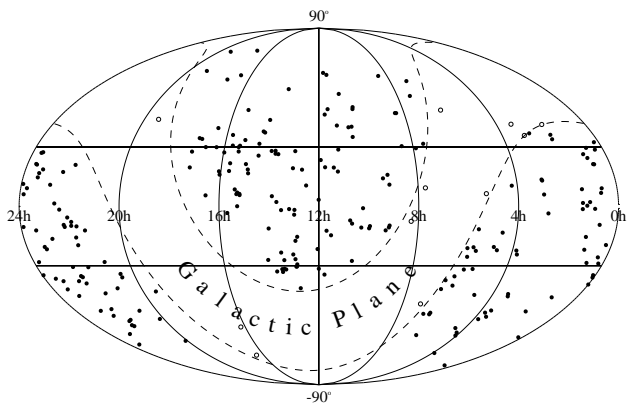}}
  \caption[]{The distribution of the 242 clusters in the statistical XBACs 
           sample on the sky (filled circles). The 11 clusters at low Galactic 
	   latitude (and $z\leq 0.2$) that 
	   have been excluded from the sample are shown as open circles.}
  \label{xbacs_skymap}
\end{figure}

\noindent
Figure~\ref{xbacs_skymap} shows the distribution of the XBACs on the sky in 
equatorial coordinates.

\section{Summary} 

\begin{figure*} 
  \epsfxsize=\textwidth
   \vspace*{2cm}
  \hspace*{2cm} \epsffile{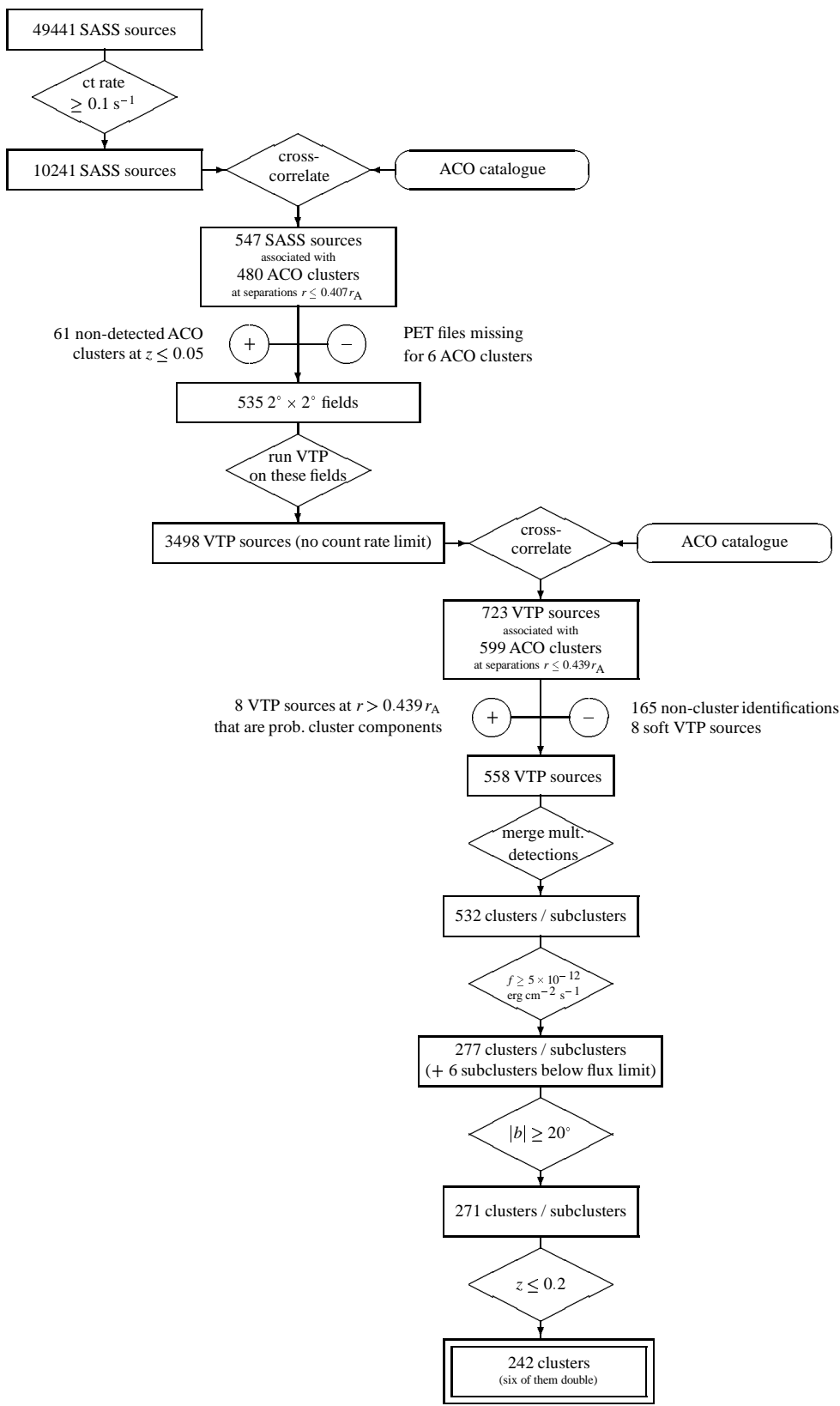} \mbox{}\\*[-3.5cm]
  \caption[]{Flow diagram summarizing the compilation of the XBACs sample}
  \label{xbacs_flow}
\end{figure*}

In this first paper of a series investigating the properties of the X-ray
brightest Abell-type clusters detected in the ROSAT All-Sky Survey (RASS) we
present the sample and describe its compilation. 

The flow diagram in Fig.~\ref{xbacs_flow} gives on overview of the 
compilation procedure, from the master list of RASS X-ray sources 
detected by the SASS to the final XBACs sample selected from VTP
detections.

The VTP re-analysis of the RASS photon data is a crucial step in the
compilation of the XBACs sample. VTP is a source detection and
characterization algorithm developed by one of us (HE) to detect
overdensities of essentially arbitrary shape in the exposure
normalized surface density of RASS photons. Since VTP makes no
assumptions about the source geometry (such as sphericity or radial
flux profile) it is particularly well suited for the detection and
characterization of extended and potentially irregular emission from
clusters of galaxies.

The VTP analysis includes a correction of the initial count rates for
low-surface brightness emission that has escaped direct
detection. This correction takes into account the PSPC point spread
function in the RASS and assumes a King law for the generalized radial
surface brightness profile. The contribution from point sources
embedded in the diffuse cluster emission is corrected for by adopting
the geometrical mean of the raw and the corrected count rate values as
the final count rate from diffuse ICM emission.  A comparison of these
final VTP values with the best count rate determinations from
archival, pointed PSPC observations for 100 selected clusters shows no
systematic differences for count rates ranging from 0.1 to 10
s$^{-1}$; the total $1\sigma$ uncertainty in the VTP count rates is
about twice as high as that from photon statistics alone.

When comparing the VTP cluster count rates with the respective SASS
values we find the SASS count rates to underestimate the true cluster
brightness by typically a factor of 0.60 (median) with the 25$^{\rm
th}$ and 75$^{\rm th}$ percentiles being 0.43 and 0.75.

Although the source extent is not used in the compilation of the
XBACs sample, we investigate its significance for future compilations
of purely X-ray selected samples. A comparison of the extent parameters
returned by the SASS and VTP shows that, at redshifts greater than 0.12,
both algorithms successfully detect some 70 per cent of our ACO clusters
as extended X-ray sources. At lower redshifts, VTP has a higher efficiency
(approaching 100 per cent at $z\sim 0.05$). If the extent information
from both algorithms are combined, essentially all XBACs are classified
as extended by at least one method irrespective of redshift.

The sample compiled from VTP detections after non-cluster sources have
been discarded consists of 277 clusters (or subclusters) above a flux
limit of completeness (95 per cent) of $5.0 \times 10^{-12}$ erg 
cm$^{-2}$ s$^{-1}$. Six more sources with X-ray fluxes below this
value are included as subclusters of systems whose brightness
exceeds the flux limit when both components are combined.

In order to remain consistent with previous studies, we limit our 
statistical sample to clusters at $|b| \geq 20^{\circ}$ and within
the nominal volume of completeness of the ACO catalogue, i.e.\ to 
redshifts lower than 0.2. This leads to our final sample of
the 242 X-ray brightest Abell-type clusters (XBACs) from the ROSAT
All-Sky Survey, 92 per cent of which have measured redshifts. The
sample's overall completeness is 80 per cent; in view of the
complicated selection procedure, we believe this figure to come close
to the maximum value achievable for such a sample. As, at its high
redshift end, our X-ray flux limited sample is intrinsically dominated
by very X-ray luminous and optically rich clusters, the volume
incompleteness in poor clusters of the underlying optical catalogue
does not affect the XBACs.

The XBACs constitute the largest X-ray flux limited sample of clusters of
galaxies compiled to date. By design, the sample is not only essentially
complete but also free from contamination by non-cluster X-ray sources. In the
following papers of this series the XBACs' properties are investigated in 
detail.

\section*{Acknowledgements} 

The authors would like to thank the ROSAT team at MPE for obtaining,
processing, and providing the RASS X-ray data this analysis is based upon, and
especially  Cristina Rosso for her tireless efforts in the RASS PET file
retrieval. The  ROSAT project is funded by the Bundesministerium f\"ur
Forschung und Technologie (BMFT) and supported by the Max-Planck-Gesellschaft
zur F\"orderung der Wissenschaften (MPG).
            
Discussions with Heinz Andernach, Sabrina De Grandi, and the referee, Pat Henry,
have helped to remove ambiguities and make this paper more readable.

Only thanks to the help of many colleagues who allowed us to use unpublished
data was it possible to ensure that 92 per cent of the XBACs now have
measured  redshifts. In particular, we are indebted to Ian Smail who allowed us
to use and explicitly quote an unpublished redshift for A\,2496 obtained by him
and one of us (ACE) at the Las Campanas 100-inch telescope. Two redshifts for
southern ACO clusters  were obtained by Claudia Mendes de Oliveira and one of
us (HE); Claudia's instrumental r\^ole in the data reduction and her permission
to publish the redshifts here are gratefully acknowledged. Usage of five
redshifts for ACO clusters in the region around the South Galactic Pole is by
kind permission from A.K.\ Romer and  collaborators (1995, private
communication). Last but not least we should like to thank Erik Tago and
Michael Ledlow for background information on redshifts referred to in
the literature.

The identification of non-cluster sources that contaminated the sample we
originally started from was greatly facilitated by the availability of
digitized optical images from the POSS and UK Schmidt sky surveys obtained
through the {\sc SkyView} facility. Thanks to Thomas McGlynn, Keith 
Scollick and co-workers for developing and maintaining {\sc SkyView}. All of
the data analysis presented in this paper was carried out using the Interactive
Data Language (IDL). We are indebted to all who contributed to the
various IDL User's Libraries; the routines of the IDL  Astronomy User's Library
(maintained by Wayne Landsman) have been used particularly extensively.

HE and ACE gratefully acknowledge financial support by European Union and PPARC
fellowships, respectively. HB thanks the BMFT for financial support from the
Verbundforschung programme 50OR93065. JPH acknowledges support by NASA Grant
NAGW-201.

Optical observations performed by one of us (JPH) were partially
obtained with the Multiple Mirror Telescope, a joint facility
of the Smithsonian Institution and the University of Arizona.
This research has made use of data obtained through the High Energy
Astrophysics Science Archive Research Center Online Service, provided by the
NASA-Goddard Space Flight Center, the Leicester Database and Archive
Service's XOBSERVER programme, the NASA/IPAC Extragalactic Database (NED),
and the SIMBAD database maintained at the Centre de Donn\'ees Astronomiques de 
Strasbourg.

\end{document}